\documentclass[superscriptaddress,aps,preprintnumbers,showpacs,prd,nofootinbib,preprint]{revtex4-1}

\usepackage{graphicx} 
\usepackage{amsmath,amssymb,amsthm,slashed,bm}
\usepackage[colorlinks,allcolors=blue]{hyperref} 

\begin{document}

%%%%%%%%%%%%%%%%%%%%%%%%%%%%%%%%%%%%%%%%%%%

\preprint{TU-1281, KEK-QUP-2025-0021}

\title{ 
Dynamical Prevention of Topological Defect Formation
}

\author{
Junseok Lee
}
\affiliation{Department of Physics, Tohoku University, 
Sendai, Miyagi 980-8578, Japan}
\author{
Kai Murai
}
\affiliation{Department of Physics, Tohoku University, 
Sendai, Miyagi 980-8578, Japan} 
\author{
Kazunori Nakayama
}
\affiliation{Department of Physics, Tohoku University, 
Sendai, Miyagi 980-8578, Japan} 
\affiliation{International Center for Quantum-field Measurement Systems for Studies of the Universe and Particles (QUP), KEK, Tsukuba, Ibaraki 305-0801, Japan}
\author{
Fuminobu Takahashi
}
\affiliation{Department of Physics, Tohoku University, 
Sendai, Miyagi 980-8578, Japan} 
\affiliation{Kavli IPMU (WPI), UTIAS, University of Tokyo, Kashiwa 277-8583, Japan}

\begin{abstract}
Topological defects can have significant cosmological consequences, so their production must be examined carefully.
It is usually assumed that topological defects are produced if the temperature becomes sufficiently high, but in reality their formation depends on the post-inflationary dynamics of a symmetry-breaking scalar.
We analyze the dynamics of a symmetry-breaking scalar field in the early universe within models that provide an effective negative mass term at the origin, and show that the symmetry can remain broken so that topological defects are never formed. In particular, we demonstrate that nonthermally produced particles (such as the Standard Model Higgs) during preheating can generate such an effective negative mass term, allowing the scalar field to follow a time-dependent minimum even in renormalizable models with a quartic coupling. We also discuss the implications of this result for the Peccei-Quinn scalar in axion models.
\end{abstract}

\maketitle
\flushbottom

\vspace{1cm}

%%%%%%%%%%%%%%%%%%%%%%%%%%%%%%%%%%%%%%
\section{Introduction}
\label{sec: intro}
%%%%%%%%%%%%%%%%%%%%%%%%%%%%%%%%%%%%%%

In many particle physics models, spontaneous symmetry breaking in the early universe can lead to the formation of topological defects, which may either cause cosmological problems or leave interesting observational signals~\cite{Vilenkin:2000jqa}.
Typical examples of topological defects are domain walls, cosmic strings, and monopoles.
If symmetry breaking occurs before inflation, such defects are inflated away, and almost none of them remain in the observable universe~\cite{Guth:1980zm}.
However, if symmetry breaking occurs after inflation, a large number of defects will populate the universe, and their cosmological implications must be carefully taken into account.
In particular, stable domain walls and monopoles quickly overclose the universe, so stringent upper bounds on the symmetry-breaking scale are imposed to avoid it.

Topological defects are usually assumed to be produced as follows.
At high temperature, thermal effects trap the Higgs field%
\footnote{
    In this paper, the scalar field that breaks global or local (gauge) symmetry is collectively referred to as ``Higgs''.
} 
at the origin.
As the universe cools, these thermal effects become negligible, and the field rolls down to the vacuum.
Since the Higgs field carries spatial fluctuations, it falls into different vacua at different locations.
Topological defects then appear since distinct vacua are not smoothly contracted if the topology is nontrivial.
If such a phase transition takes place after inflation, it is expected that a large number of defects will emerge in the universe.

However, the dynamics of the Higgs field in the actual universe need not be so simple.
For example, if the Higgs field couples to the Ricci curvature with an appropriate sign, it may acquire a negative Hubble mass term during inflation and develop a large field value.
In this case, the symmetry is already broken during inflation, so the defects are inflated away.
Nevertheless, depending on the post-inflationary dynamics of the scalar field, defects may be regenerated.
Let us consider a simple model with the following scalar potential:
\begin{align}
    V = -(m^2 + c H^2) |S|^2 + \lambda |S|^4,
\end{align}
where $S$ is the Higgs field, $H$ is the Hubble parameter, $m$ is the mass parameter, and $c$ and $\lambda$ are numerical constants assumed to be positive.
A prominent example is the Peccei-Quinn (PQ) field to address the strong CP problem~\cite{Peccei:1977hh,Peccei:1977ur}, which admits stable string-domain wall networks after the U(1) and QCD phase transition depending on axion models~\cite{Kawasaki:2013ae,Marsh:2015xka,DiLuzio:2020wdo}.
If $m$ is much larger than the inflationary Hubble scale as well as the maximum temperature after inflation, the Higgs field 
remains in the broken phase,
and the symmetry is never restored. 
This case is out of our interest, so throughout this paper we consider the case in which $m$ is negligibly small compared with the inflationary Hubble scale.
As will be discussed in the next section, while the Higgs field indeed acquires a large field value during inflation in this case, after inflation ends, the field starts oscillating around the origin, $S=0$.
It may lead to the formation of topological defects through particle production and the resulting thermal/nonthermal symmetry restoration~\cite{Kofman:1995fi,Tkachev:1995md,Kasuya:1998td,Tkachev:1998dc}
(see Refs.~\cite{Kasuya:1996ns,Moroi:2013tea,Kawasaki:2013iha,Graham:2025iwx} for the case of the PQ scalar).
This example illustrates how crucial it is to properly take into account the dynamics of the Higgs field throughout the history of the universe.

In this paper, we follow the dynamics of the Higgs field from inflation through the post-inflationary stage and derive the condition under which the Higgs field never crosses the origin. This is a \textit{sufficient} condition for preventing defect formation.
As a starting point, we consider a simple model in which the Higgs field attains a large value due to a negative Hubble-induced mass term. The potential is assumed to be%
\footnote{
    This type of potential is often considered in the context of Affleck-Dine baryogenesis (negative $m^2$ in this case)~\cite{Dine:1995uk,Dine:1995kz} 
    and PQ scalar stabilization models~\cite{Murayama:1992dj,Banks:2002sd,Choi:2011rs,Nakayama:2012zc}
    in supersymmetric (SUSY) theories.
    The appearance of the Hubble mass term is ubiquitous even in non-SUSY theories since we can always introduce the Higgs coupling to the Ricci curvature $R$ as $\mathcal L \sim R|S|^2$. 
}
\begin{align}
    V = -(m^2 + c H^2) |S|^2 + \lambda |S|^n.
    \label{eq:VH}
\end{align}
We will obtain a lower bound on $n$ to avoid  defect formation, as already pointed out in Ref.~\cite{Harigaya:2015hha}.
For the simplest renormalizable potential $n=4$, the zero-crossing of $S$ is unavoidable.
Thus we need another ingredient to avoid the zero-crossing for the most interesting case $n=4$.
To this end, we introduce a coupling of $S$ to another light (real) scalar field $\chi$ as
\footnote{
    This scalar field $\chi$ may be identified with the Standard Model Higgs boson (taking account of the four real degrees of freedom), though we do not specify it here.
}
\begin{align}
    V = -(m^2 + c H^2 + \lambda_{S}\chi^2) |S|^2 + \lambda |S|^4,
    \label{eq:Vnegative}
\end{align}
where $\lambda_{S}$ is a positive constant.
We assume that $\chi$ does not acquire a vacuum expectation value, which will be  justified later.
If $\chi$ is in the thermal bath, this coupling gives a negative thermal mass to the Higgs field $S$, which tends to keep $S$ away from the origin and avoid the zero-crossing.
We will show that,
for sufficiently high temperatures, $S$ does not cross the origin in this case,
and hence the topological defect formation is avoided.
However, this situation may not be realistic, since immediately after inflation $\chi$ particles are considered to be far from thermal equilibrium on the time scale of a few inflaton oscillations. 
Thus we will also consider a concrete setup in which the inflaton couples to $\chi$ and $\chi$ particles are produced through the preheating process~\cite{Kofman:1994rk,Kofman:1997yn}.
We will see the effect of (non-equilibrium) $\chi$ particle production on the $S$ dynamics and derive the condition to avoid the zero-crossing.

This paper is organized as follows. 
In Sec.~\ref{sec:Hub}, we analyze the Higgs dynamics in the model (\ref{eq:VH}).
In Sec.~\ref{sec:ther}, we analyze the Higgs dynamics in the model (\ref{eq:Vnegative}), assuming that $\chi$ is instantaneously thermalized after inflation.
In Sec.~\ref{sec:noneq}, we consider the same model (\ref{eq:Vnegative}), but here $\chi$ particles are produced through preheating and have (highly) non-equilibrium distribution.
We discuss the implications of our results for axion models in Sec.~\ref{sec:PQ}.
We conclude in Sec.~\ref{sec:conc}.

%%%%%%%%%%%%%%%%%%%%%%%%%%%%%%%%%%%%%%
\section{Higgs dynamics with a negative Hubble mass term}
\label{sec:Hub}
%%%%%%%%%%%%%%%%%%%%%%%%%%%%%%%%%%%%%%

We consider the cosmological evolution of a Higgs field $S$ which is in the representation of the group $G$, e.g., $G = {\mathbb Z}_2$, U(1), SU(2).
If $S$ has a vacuum expectation value that is nontrivial under transformations, $S \rightarrow U^{-1} S U$, where $U$ is a unitary operator of $G$ on the Hilbert space, the symmetry is spontaneously broken.
Topological defects would be formed and persist if there is a period when such a symmetry is dynamically restored in the cosmological history after inflation.

%%%%%%%%%%%%%%%%%%%%%%%%%%%%%%%%%%%%%%
\subsection{Symmetry (non-)restoration through the Hubble mass}
\label{subsec: SymRes}
%%%%%%%%%%%%%%%%%%%%%%%%%%%%%%%%%%%%%%

First, let us examine the dynamics of the Higgs with a negative Hubble mass term.  The potential is given by
\begin{equation}
    V = - c H^2 |S|^2 + \frac{2^{n/2}}{n}\lambda_n |S|^n,
\end{equation}
where $n$ is an even integer larger than 2, and $\lambda_n$ is a positive self-coupling constant.
The bare mass $m$ of $S$ in Eq.~(\ref{eq:VH}) is assumed to satisfy $m \ll H_{\rm inf}$.
Since our main focus is on the post-inflationary dynamics shortly after inflation, when the Hubble parameter is still of order $H_{\rm inf}$, we neglect the bare mass term in what follows.
We are interested in the case of positive $c$, which leads to the spontaneous symmetry breaking.
Since only the radial component of $S$ is relevant, we can further simplify the dynamics by concentrating on the radial component as
\begin{equation}
    |S| = \frac{\sigma}{\sqrt{2}}.
\end{equation}
Then the potential becomes
\begin{equation}
    V_\sigma = - \frac{1}{2} c H^2 \sigma^2 + \frac{1}{n}\lambda_n \sigma^n .
\end{equation}
The time-dependent potential minimum is located at
\begin{equation}
    \sigma_{\rm min} = \left( \frac{c H^2}{\lambda_n} \right)^{\frac{1}{n-2}},
    \label{eq:sigmamin}
\end{equation}
around which the mass squared for $\sigma$ is ${(n-2)} c H^2$.
If $c \gtrsim \mathcal O(1)$, $\sigma$ settles at the potential minimum $\sigma_{\rm min}$, and hence the symmetry is broken during inflation. Thus, any topological defects that might have formed during inflation are inflated away.
Below we qualitatively estimate the dynamics of $\sigma$ after inflation ends to examine whether regeneration of topological defects occurs or not.

For simplicity, in this paper we assume that $S$ is homogeneous and settles down to the potential minimum at the initial time.
In fact, in addition to the radial degree of freedom, the direction corresponding to the broken symmetry, which is physically eaten by the gauge field if the symmetry is gauged, should also be considered when the broken symmetry is continuous.
Even if $\sigma$ does not overshoot the symmetry restoration point, amplification of angular fluctuations through a momentum transfer from the radial component may still result in the non-zero winding number.
However such non-thermal restoration is considered not to take place: the fluctuations in the angular component do not reach the symmetry breaking scale and decrease after parametric resonance ends, as shown in Refs.~\cite{Ema:2017krp,Kawasaki:2017kkr}.
Thus an analysis based solely on a homogeneous mode is satisfactory for providing a sufficient condition for preventing defect formation.

The zero mode of $\sigma$ follows the equation of motion,
\begin{equation}
    \ddot{\sigma} + 3H\dot{\sigma} + \frac{\partial V_{\sigma}}{\partial \sigma} = 0 .
    \label{eq:sigmaEoM}
\end{equation}
While $\sigma$ tends to asymptotically reach the temporary minimum due to the Hubble mass, it is not obvious that $\sigma$ can fully trace the evolution of the potential minimum, especially soon after the inflation ends.
Here, we assume $c \gtrsim \mathcal{O}(1)$ so that $\sigma$ oscillates around $\sigma_\mathrm{min}$.
Due to the cosmic expansion, the oscillation amplitude of $\sigma$ around $\sigma_\mathrm{min}$, which we denote by $\delta \sigma$, decreases in time.
However, if it decreases more slowly than $\sigma_\mathrm{min}$, $\sigma$ could eventually overshoot the origin.
If the rate $|\dot{\sigma}_{\rm min}/\sigma_{\rm min}|$ is sufficiently slower than the mass scale,
\begin{equation}
    \sqrt{(n-2)c}\,H 
    >
    \left| \frac{\dot{\sigma}_{\rm min}}{\sigma_{\rm min}} \right|
    = \frac{2}{n-2} \frac{|\dot{H}|}{H}
    ,
\end{equation}
which means $c>9(1+w)^2/(n-2)^3$, where $w$ is the equation of state parameter such that $H^2 \propto a^{-3(1+w)}$, the change is adiabatic and the occupation number of $\sigma$ is conserved.
Assuming the adiabatic change, the amplitude of the oscillation around the potential minimum, $\delta\sigma$, decreases as%
\footnote{
    Some amount of oscillation of $\sigma$ around its temporary minimum $\sigma_{\rm min}$ is inevitably induced at the end of inflation~\cite{Nakayama:2011wqa}.
}
\begin{equation}
    \delta\sigma \propto a^{-\frac{3}{4}(1-w)} ,
    \label{eq:delta sigma}
\end{equation}
since the comoving number density should be conserved: $H\,\delta\sigma^2 \propto a^{-3}$ (see App.~\ref{app:cons}).
Here, we used the fact that the mass of $\delta \sigma$ is given by $\sqrt{(n-2)c} H$.
If the ratio between the amplitude and the distance between the minimum and the origin ($\sigma=0$),
\begin{equation}
    \frac{\delta\sigma}{\sigma_{\rm min}}
    \propto
    a^{-\frac{3(n-6)}{4(n-2)}+\frac{3(n+2)}{4(n-2)}w}
    ,
\end{equation}
is growing with time, $\sigma$ could overshoot the origin.
The exponent is negative if
\begin{equation}
    n > n_{\rm crit} \equiv \frac{2(3+w)}{1-w}
    \quad
    {\rm or}
    \quad
    w < \frac{n-6}{n+2},
    \label{eq:nbound}
\end{equation}
where we used $n>2$ and assumed $w < 1$.
Therefore, requiring that $\sigma$ keeps track of the temporary minimum of the potential leads to a lower bound on $n$ as $n>6$ for $w=0$ and $n>10$ for $w=1/3$, for example.%
\footnote{
    The dynamics in the critical case $n=n_{\rm crit}$ is highly nontrivial and requires a separate analysis~\cite{Ema:2015dza}.
Also, the analysis of the temperature-dependent non-renormalizable term is given in Appendix~\ref{app:nonren}.}
The same result is obtained by directly analyzing the equation of motion after an elaborate change of variables~\cite{Dine:1995kz,Harigaya:2015hha}.%
\footnote{
If $c$ is very large, the initial oscillation amplitude $\delta \sigma$ is suppressed.
If the Hubble mass term becomes comparable to the bare mass term before $\delta \sigma/\sigma_\mathrm{min}$ becomes $\mathcal{O}(1)$, the zero-crossing will not take place. 
In this work, we do not consider such a large $c$.
}

%%%%%%%%%%%%%%%%%%%%%%%%%%%%%%%%%%%%%%
\subsection{Numerical results}
\label{subsec: Inflation}
%%%%%%%%%%%%%%%%%%%%%%%%%%%%%%%%%%%%%%

In order to capture the precise dynamics of $\sigma$, we need to know the precise time evolution of the Hubble parameter, which in turn depends on the inflaton dynamics.
The background motion of the inflaton $\phi$ is given by solving the equation of motion,
\begin{equation}
    \ddot{\phi} + 3H\dot{\phi} + \frac{\partial V_{\rm inf}}{\partial \phi} = 0,
\end{equation}
where $H = \frac{1}{\sqrt{3} M_{\rm pl}}\sqrt{\frac{1}{2}\dot{\phi}^2 + V_{\rm inf}}$ is the Hubble parameter with the reduced Planck mass $M_{\rm pl} = (8\pi G)^{-1/2}$, and $V_{\rm inf}$ is the potential for the inflaton.
First let us assume a quadratic chaotic inflation model~\cite{Linde:1983gd}:
\begin{align}
    V_{\rm inf}(\phi) = \frac{1}{2}m_\phi^2\phi^2
\end{align}
where $m_\phi$ is the inflaton mass.
Inflation ends around $\phi = \sqrt{2}\, M_{\rm pl}$ where the slow-roll parameters become $\mathcal O(1)$, and the initial field value corresponding to the e-folds ${\cal N} = 50$ from the end of slow-roll inflation is given by $\phi \simeq 10\sqrt{2}\, M_{\rm pl}$.
The Hubble parameter during the inflationary period is given by $H \simeq \frac{1}{\sqrt{6}} m_\phi \phi / M_{\rm pl}$.

\begin{figure}[!t]
\begin{center}  
\includegraphics[width=0.49\textwidth]{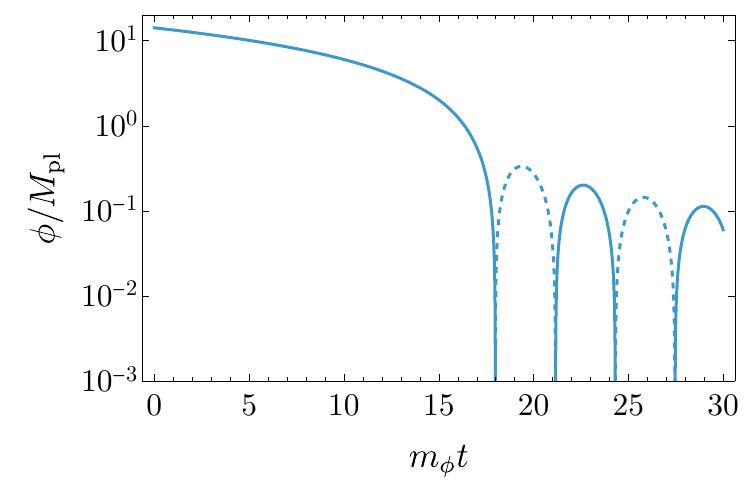}
\includegraphics[width=0.49\textwidth]{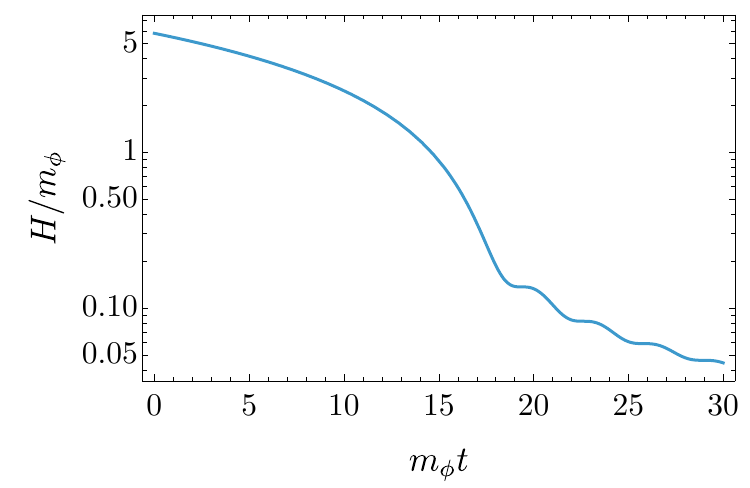}
\end{center}
\caption{%
    The time evolution of the inflaton field (left) and the Hubble parameter (right) in the chaotic inflation model.
    The dashed line in the left panel corresponds to the negative $\phi$.
    }
\label{fig: inflaton}
\end{figure}

The left panel of Fig.~\ref{fig: inflaton} shows the time evolution of the inflaton field.
The inflaton field is initially set to be $\phi \simeq 10\sqrt{2}\, M_{\rm pl}$ at $t = 0$, and the slow-roll phase ends at $m_\phi t \simeq 16$, after which $\phi$ starts to oscillate.
The right panel of Fig.~\ref{fig: inflaton} shows the time evolution of the Hubble parameter.
%%%%%%%%%%%%%%%%%%%%%%%%%%%%%%%%%%%%%%
%%%%%%%%%%%%%%%%%%%%%%%%%%%%%%%%%%%%%%

\begin{figure}[!t]
\begin{center}
    \includegraphics[width=0.49\textwidth]{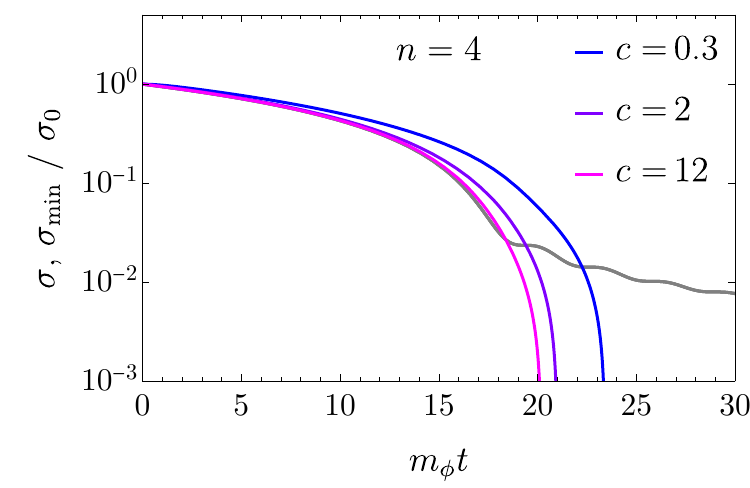}
    \includegraphics[width=0.49\textwidth]{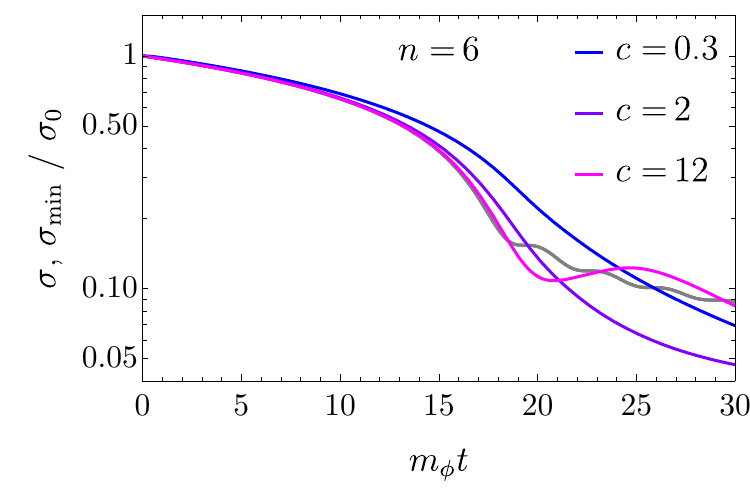}
    \includegraphics[width=0.49\textwidth]{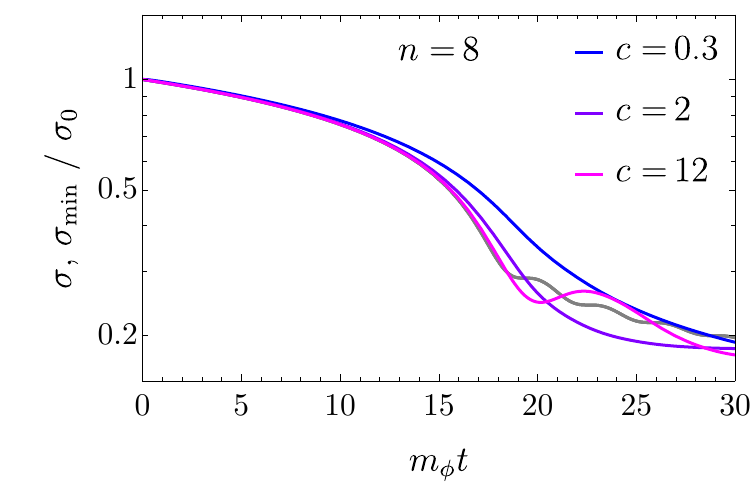}
\end{center}
\caption{%
    The numerical results of the motion of $\sigma$ (colored lines) normalized by $\sigma_0$ in the chaotic inflation model.
    Each panel shows a different $n$ case, and the colors of lines differ by the coupling constant $c$.
    The gray lines show the time evolution of $\sigma_{\rm min}$, the potential minimum, which is also normalized by $\sigma_0$.
    We fix the self-coupling constant of $\sigma$ by $\lambda_n = m_\phi^{4-n}$ for $n = 6,8$.
    Since the quartic coupling constant can be scaled out through a field redefinition, $\sigma \rightarrow \sqrt{\lambda_4} \sigma$, in the classical level, $\sigma(t) / \sigma_0$ does not depend on $\lambda_4$.
}
\label{fig: sigmaHM}
\end{figure}

Given the result of the Hubble parameter as shown in the right panel of Fig.~\ref{fig: inflaton}, the motion of $\sigma$ is obtained by solving the equation of motion (\ref{eq:sigmaEoM}).
The initial value of $\sigma$ and its velocity, $\sigma_0$ and $\dot{\sigma}_0$, are set to be the potential minimum at the initial time and its velocity, respectively:
\begin{align}
    \sigma_0 &= \sigma_{\rm min} |_{t=0}, \\
    \dot{\sigma}_0 &= \dot{\sigma}_{\rm min} |_{t=0}.
\end{align}
The numerical results are shown in Fig.~\ref{fig: sigmaHM}.
The gray lines show the potential minimum $\sigma_{\rm min}$ (\ref{eq:sigmamin}), and the colored lines show the time evolution of $\sigma$, all of which are normalized by the initial value $\sigma_0$.
In the case of $n=4$, it is observed that $\sigma$ crosses the origin $\sigma=0$ within a few oscillations of the inflaton, whereas this does not occur for higher $n \geq 6$.
This is consistent with the lower bound on $n$ (\ref{eq:nbound}) by substituting $w=0$, since the inflaton oscillation in this model behaves as non-relativistic matter.\footnote{
    In fact, this is nontrivial because the equation-of-state parameter during inflaton oscillations is not exactly zero, especially immediately after the end of inflation.
}
Here and hereafter, we do not follow the dynamics of $\sigma$ after it reaches zero because the following dynamics is not fully captured by the simplified setup of the real degree of freedom $\sigma$.
Note that, for $n\geq n_{\rm crit}$, $\sigma$ does not cross the origin even for small value of $c$, or even without the negative Hubble mass term for appropriate initial conditions. This is because the scalar with $n>n_{\rm crit}$ follows the so-called scaling solution without oscillation~\cite{Liddle:1998xm,Dimopoulos:2003ss}. (See Ref.~\cite{Ema:2015dza} for the case of $n=n_{\rm crit}$.)

%%%%%%%%%%%%%%%%%%%%%%%%%%%%%%%%%%%%%%
%%%%%%%%%%%%%%%%%%%%%%%%%%%%%%%%%%%%%%

Next let us consider a new (or hilltop) inflation model~\cite{Linde:1981mu,Albrecht:1982wi,Boubekeur:2005zm}: 
\begin{align}
    V_{\rm inf}(\phi) = \Lambda^4 \left[1-\left(\frac{\phi}{v}\right)^\ell\right]^2,
    \label{eq:NewInf}
\end{align}
where $\Lambda$ and $v$ are parameters with mass dimension, and $\ell$ is an integer.%
\footnote{
    This potential is naturally obtained in SUSY~\cite{Kumekawa:1994gx,Izawa:1996dv,Asaka:1999yd,Asaka:1999jb,Senoguz:2004ky,Nakayama:2012dw}. Also, a similar hilltop inflation is realized in the multi-natural inflation~\cite{Czerny:2014wza,Czerny:2014xja}.
}
In this model inflation happens around $\phi=0$, and the slow-roll phase ends around
\begin{align}
	\phi_{\rm end} = v \left[ \frac{1}{2\ell(\ell-1)} \frac{v^2}{M_{\rm pl}^2} \right]^{1/(\ell-2)}.
\end{align}
The typical Hubble parameter during inflation is given by $H^2_{\rm inf} \simeq \Lambda^4 / (3 M_{\rm pl}^2)$.
In contrast to the chaotic inflation, generally there is a hierarchy between the Hubble parameter and the inflaton mass at the potential minimum: $m_\phi = \sqrt{6} \ell M_{\rm pl} H_{\rm inf} / v$.

\begin{figure}[!t]
\begin{center}  
\includegraphics[width=0.49\textwidth]{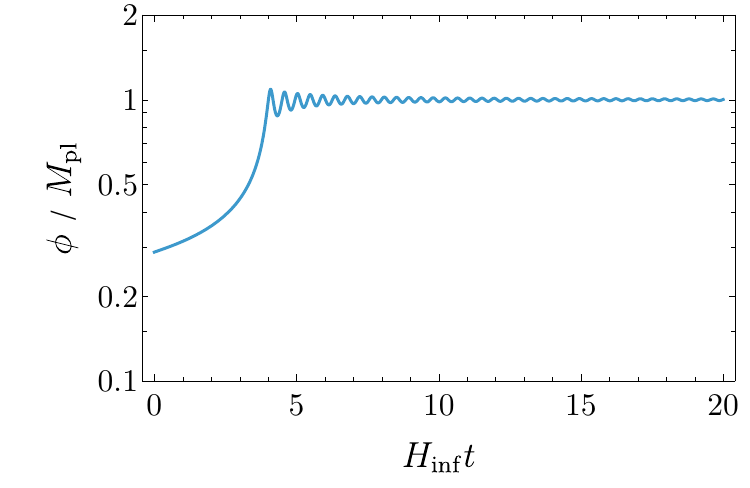}
\includegraphics[width=0.49\textwidth]{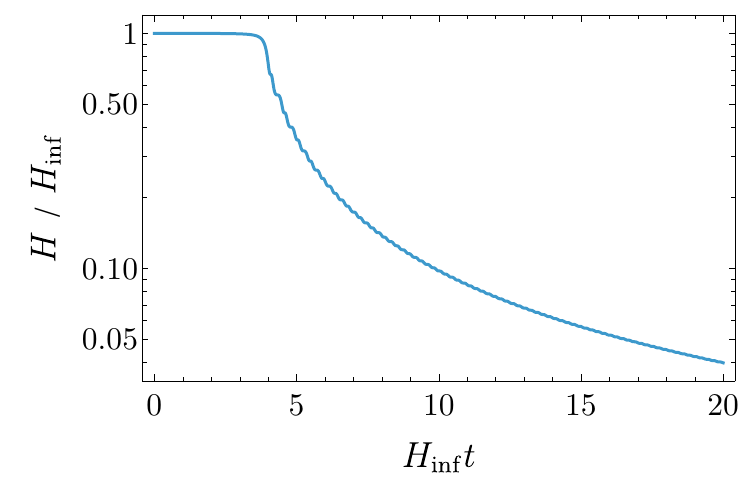}
\end{center}
\caption{%
    The time evolution of the inflaton field (left) and the Hubble parameter (right) in the new inflation model (\ref{eq:NewInf}) for $\ell=6$ and $v=M_{\rm pl}$. The time is normalized by the Hubble parameter during inflation, $H_{\rm inf}$.
}
\label{fig: inflaton_NI}
\end{figure}

\begin{figure}[!t]
\begin{center}
    \includegraphics[width=0.49\textwidth]{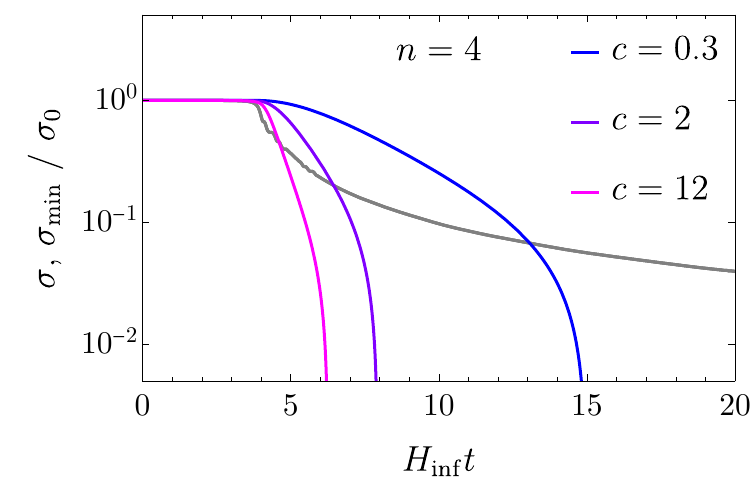}
    \includegraphics[width=0.49\textwidth]{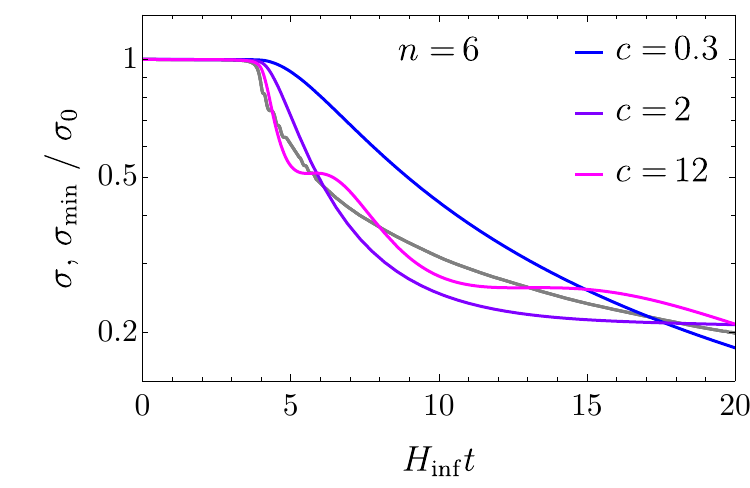}
    \includegraphics[width=0.49\textwidth]{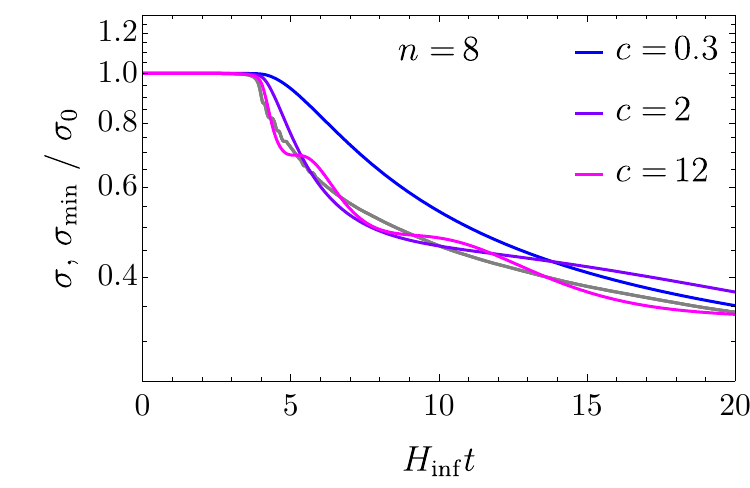}
\end{center}
\caption{%
    Same as Fig.~\ref{fig: sigmaHM} but for the new inflation model (\ref{eq:NewInf}). The time is normalized by the Hubble parameter during inflation, $H_{\rm inf}$.
}
\label{fig: sigmaHM_NI}
\end{figure}

Fig.~\ref{fig: inflaton_NI} shows the time evolution of the inflaton (left panel) and the Hubble parameter (right panel).
We choose $\ell=6$ and $v=M_{\rm pl}$ with the initial condition $\phi = 0.8\, \phi_{\rm end}$.
The evolution of $\sigma$ for different values of $n$ and $c$ are shown in Fig.~\ref{fig: sigmaHM_NI}.
As in the chaotic inflation case, 
for $n = 4$ the field
$\sigma$ overshoots the origin  within several Hubble times,  whereas
for $n\geq 6$ it does not.
This confirms that the lower bound on $n$ (\ref{eq:nbound}) to avoid the zero-crossing is insensitive to the choice of inflation models.
An obvious difference between chaotic and new inflation models is that, for the new inflation model, the inflaton oscillates many times before $\sigma$ crosses the origin due to the hierarchy between the inflaton mass and the Hubble scale.
This feature has an impact on the dynamics of the model with a negative Higgs coupling considered in the following sections.

%%%%%%%%%%%%%%%%%%%%%%%%%%%%%%%%%%%%%%
%%%%%%%%%%%%%%%%%%%%%%%%%%%%%%%%%%%%%%

%%%%%%%%%%%%%%%%%%%%%%%%%%%%%%%%%%%%%%
%%%%%%%%%%%%%%%%%%%%%%%%%%%%%%%%%%%%%%

%%%%%%%%%%%%%%%%%%%%%%%%%%%%%%%%%%%%%%
\section{Higgs dynamics with a negative coupling: thermal case}
\label{sec:ther}
%%%%%%%%%%%%%%%%%%%%%%%%%%%%%%%%%%%%%%

We have shown in the previous section that $\sigma$ inevitably crosses the origin in the minimal renormalizable setup with $n=4$.
In this section, we show that such a zero crossing can be avoided by introducing a negative portal coupling to a thermalized field such as Eq.~(\ref{eq:Vnegative}):
\begin{align}
    V = -\frac{1}{2}(c H^2 + \lambda_{\chi\sigma}\chi^2) \sigma^2 + \frac{\lambda_4}{4} \sigma^4,
    \label{eq:chisigma}
\end{align}
We will focus on the case $n=4$ in the following.

The oscillation of the inflaton after inflation drives the reheating process.
We assume that the inflaton perturbatively decays into $\chi$ particles after the end of inflation and that $\chi$ particles are thermalized instantaneously.
The coupling to thermalized $\chi$ provides a thermal mass for $\sigma$:
\begin{align}
    V_T = -\frac{1}{2}\left(c H^2 + \frac{\lambda_{\chi\sigma}}{12} T^2\right) \sigma^2 + \frac{\lambda_4}{4} \sigma^4.
\end{align}
The thermal contribution is negative, which keeps $\sigma$ away from the origin during the post-inflationary evolution.
For  sufficiently high temperatures, $T \gg H$, the thermal mass term dominates, and the potential minimum is
\begin{align}
    \sigma_{{\rm min},T} = \sqrt{\frac{\lambda_{\chi\sigma}}{12 \lambda_4}} T,
\end{align}
around which the mass squared for $\sigma$ is $\lambda_{\chi\sigma} T^2 / 6$.
Thus, $\sigma$ is strongly stabilized at this temperature-dependent minimum.\footnote{
    This situation is sometimes called ``inverse symmetry breaking'', since the symmetry breaking happens at high temperature~\cite{Weinberg:1974hy,Dvali:1995cc}. 
    See also Refs.~\cite{Jansen:1998rj,Pinto:1999pg} for lattice studies.
    This idea may also be applied to the electroweak symmetry breaking~\cite{Meade:2018saz,Baldes:2018nel} and the PQ symmetry breaking~\cite{Dvali:1995cc,Ramazanov:2022kbd}.
}
Actually the ratio of the oscillation amplitude of $\sigma$ around the temporary minimum $\delta\sigma$ and $ \sigma_{{\rm min},T}$ scales as~\cite{Murai:2024alz}
\begin{align}
    \frac{\delta\sigma}{\sigma_{{\rm min},T}}\propto \begin{cases}
        a^{-15/16} & (T>T_{\rm R}) \\
        {\rm const} & (T<T_{\rm R})
    \end{cases},
\end{align}
where $T_{\rm R}$ denotes the reheating temperature and assumed matter domination for $T>T_{\rm R}$.
Therefore, before the completion of the reheating, the relative amplitude decreases and the Higgs field is safely stabilized at the temperature-dependent minimum thereafter.

\begin{figure}[!t]
\begin{center}
    \includegraphics[width=0.49\textwidth]{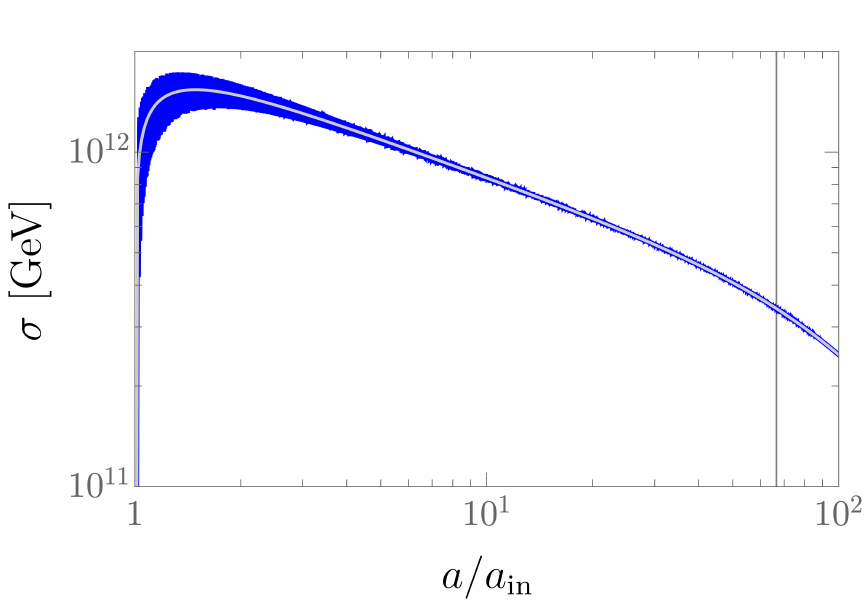}
    \includegraphics[width=0.49\textwidth]{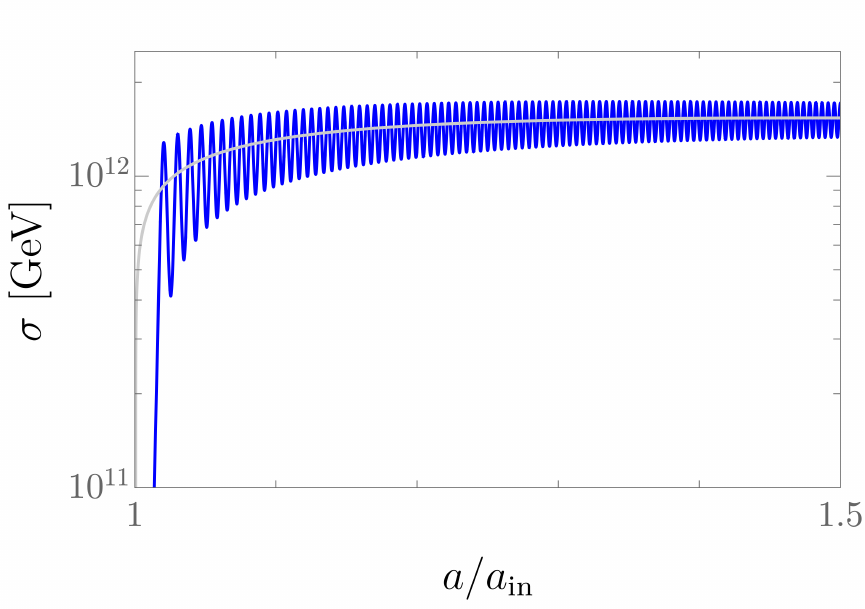}
\end{center}
\caption{%
    The numerical results of the evolution of $\sigma$ (blue lines) in the thermal case.
    The horizontal axis represents the scale factor $a$ normalized to the initial value $a_{\rm in}$.
    The right panel shows a close-up of the left panel.
    The gray lines show the time evolution of $\sigma_{\rm min}$.
    The vertical line in the left panel denotes $H = H_{\rm R}$.
    We used $c = 1$, $\lambda_{\chi \sigma} = 0.5$, $\lambda_4 = 0.3$,
    $H_\mathrm{inf} = 10^9$\,GeV, and $T_{\rm R} = 10^{12}$\,GeV.
}
\label{fig: sigmaThermal}
\end{figure}

To confirm this, we numerically solve the dynamics.
The relevant equations are
\begin{align}
    &3 H^2 M_{\rm pl}^2 = \rho_\phi + \rho_r,\\
    &\dot\rho_\phi + (3H + \Gamma_\phi)\rho_\phi = 0,\\
    &\dot\rho_r + 4H\rho_r = \Gamma_\phi\rho_\phi,\\
    &\ddot\sigma + 3H\dot\sigma + \frac{\partial V_T}{\partial\sigma} = 0,
\end{align}
where $\rho_\phi$ is the inflaton energy density, $\Gamma_\phi$ is the inflaton decay rate taken to be constant, and the radiation energy density $\rho_r$ is related to the temperature $T$ through $\rho_r = (\pi^2 g_*/30) T^4$, since we assume that decay products are instantaneously thermalized.
Here, we define the reheating temperature by $\rho_r(T_{\rm R}) = 3 H_{\rm R}^2 M_{\rm pl}^2$ with $3 H_{\rm R} = \Gamma_\phi$.
We set the initial conditions by
\begin{align}
    \rho_\phi(t_\mathrm{in}) = 3 H_\mathrm{inf}^2 M_\mathrm{pl}^2
    \ , \quad
    \rho_r(t_\mathrm{in}) = 0
    \ , \quad
    \sigma(t_\mathrm{in}) = \sqrt{\frac{c}{\lambda_4}} H_\mathrm{inf}
    \ .
\end{align}
We show the numerical solution for $\sigma$ in Fig.~\ref{fig: sigmaThermal}. 
Here, we used $g_* = 106.75$, $H_\mathrm{inf} = 10^9$\,GeV, and $T_{\rm R} = 10^{12}$\,GeV.
The numerical solution (blue) oscillates around the temporary potential minimum (gray) with a decreasing amplitude and never crosses the origin.

%%%%%%%%%%%%%%%%%%%%%%%%%%%%%%%%%%%%%%
%%%%%%%%%%%%%%%%%%%%%%%%%%%%%%%%%%%%%%

Before closing the section, we make comments on the condition for the potential so that the above discussion holds.
While we did not specify the potential for $\chi$, the total potential of $\sigma$ and $\chi$ should be bounded from below.
To this end, we introduce a quartic potential of $\chi$,
\begin{align}
    V_\chi
    =
    \frac{\lambda_\chi}{4} \chi^4
    \ ,
\end{align}
where we assume $\lambda_\chi$ satisfies $\lambda_4 \lambda_\chi > \lambda_{\chi \sigma}^2$.
First, we consider the backreaction of $\sigma$ on $\chi$.
When $\sigma$ settles to $\sigma_{\mathrm{min},T}$, it gives 
\begin{align}
    V_{\chi,T}
    =
    \frac{1}{2}\left( \frac{\lambda_\chi}{4} - \frac{\lambda_{\chi \sigma}^2}{12 \lambda_4} \right) T^2 \chi^2
    \ .
\end{align}
While $\chi$ may receive effective potentials from its couplings to other particles, we neglect such contributions here.
Then, this potential gives a positive mass term for $\chi$.
Moreover, as long as $\lambda_\chi \lesssim 1$, this thermal mass is smaller than $T$, and thus the thermal distribution of $\chi$ is not suppressed by its effective mass.

In addition, we have to consider the production of $\sigma$ from thermal $\chi$ particles.
If $\sigma$ is efficiently produced from $\chi$ via the $\lambda_{\chi\sigma}$ interaction and thermalizes, the typical amplitude of $\sigma$ will be comparable to that of $\chi$ and is given by $\delta \sigma_T \sim T/\sqrt{12}$.
To keep $\sigma$ away from the origin, we require $\delta \sigma_T \lesssim \sigma_{\mathrm{min},T}$, leading to $\lambda_{\chi \sigma} \gtrsim \lambda_4$.
Consequently, the possible range of $\lambda_4$ is limited as $\lambda_{\chi \sigma}^2 \lesssim \lambda_4 \lesssim \lambda_{\chi \sigma}$ for $\lambda_\chi \sim 1$.
The parameters used in Fig.~\ref{fig: sigmaThermal} satisfy this condition.%
\footnote{
If we identify $\chi$ with the Standard Model Higgs, we need $\lambda_{\chi \sigma}^2 / 0.1 \lesssim \lambda_4 \lesssim \lambda_{\chi \sigma}$.
}

%%%%%%%%%%%%%%%%%%%%%%%%%%%%%%%%%%%%%%
\section{Higgs dynamics with a negative coupling: non-equilibrium case}
\label{sec:noneq}
%%%%%%%%%%%%%%%%%%%%%%%%%%%%%%%%%%%%%%

In the previous section we have shown that the negative thermal mass can keep the Higgs field $\sigma$ away from the origin throughout the history of the universe.
However, the assumption that $\chi$ particles are instantaneously thermalized is not always justified.
In particular, for high-scale inflation such as chaotic inflation,
the time scale of the inflaton dynamics $(\sim \mathcal O(m_\phi^{-1}))$ is comparable to the Hubble time $H^{-1}$ just after inflation, which corresponds to the time scale of Higgs dynamics without finite density effect.
Since the thermalization time scale of $\chi$ is typically much longer than the inflaton oscillation time scale, $\chi$ particles are not fully thermalized within the time scale of the Higgs dynamics.\footnote{
    For low-scale inflation models such as new inflation, on the other hand, the inflaton oscillates many times within one Hubble time and hence the use of perturbative decay rate $\Gamma_\phi$ and the instant thermalization, as done in the previous section, can be justified.
}
In this section we consider a more realistic situation where $\chi$ particles are produced through the preheating process and have a highly non-equilibrium distribution, and we will see whether such particles can prevent the zero-crossing of the Higgs or not.

%%%%%%%%%%%%%%%%%%%%%%%%%%%%%%%%%%%%%%
\subsection{Preheating}
\label{subsec:NumRad}
%%%%%%%%%%%%%%%%%%%%%%%%%%%%%%%%%%%%%%

The period of our interest is soon after inflation when the inflaton has oscillated only a few times and thus the reheated field is not yet fully thermalized.
During this stage, the precise evolution of $\sigma$ highly depends on the dynamics of $\chi$.
Below we assume chaotic inflation with $V_{\rm inf}=m_\phi^2\phi^2/2$.
Let us suppose that a scalar field $\chi$ couples to the inflaton through
\begin{equation}
    {\cal L} \supset -\frac{1}{2} g^2 \phi^2 \chi^2 ,
\end{equation}
where $g$ is a coupling constant. Through this coupling, $\chi$ acquires a time-varying effective mass in the oscillating inflaton background, resulting in the particle creation of $\chi$.
Note that $g$ is taken to make the typical value of $g\phi$ larger than the Hubble parameter $H$.
For simplicity, we neglect self-interaction of $\chi$ as well as possible interactions with Standard Model particles. If such interactions were present, they would only accelerate thermalization and hence make the situation closer to the case of the negative thermal mass discussion in the previous section.%
\footnote{
    Actually, the Standard Model Higgs boson is always a natural candidate for $\chi$ in any concrete model, and it has sizable gauge and Yukawa couplings to the Standard Model particles.
} 

The equation of motion of $\chi$ is
\begin{equation}
    \ddot{\chi} + 3H \dot{\chi} - \frac{1}{a^2} \vec{\nabla}^2 \chi + g^2 \phi^2 \chi = 0 .
    \label{eq:chiEoM}
\end{equation}
Here we neglect the coupling to $\sigma$ as in Eq.~(\ref{eq:Vnegative}).
This can be justified since the typical value of the inflaton field is of order $M_{\rm pl}$, which is much larger than $\sigma_{\rm min}$, the typical value of $\sigma$, as we discuss in detail later.
After the coupling to $\chi$ dominates the negative mass term of $\sigma$, the potential minimum can shift to the order of $\sigma_{\rm min} \sim \sqrt{\lambda_{\chi\sigma} / 12 \lambda_4} T_{\rm R}$, which can also be neglected unless $g$ is extremely small. Here $\chi$ is assumed to be in thermal equilibrium.

We calculate the production of $\chi$ following Ref.~\cite{Kofman:1997yn}.%
\footnote{
    Our methodology is similar to Ref.~\cite{Hagihara:2018uix}, in which the modulus dynamics induced by the particles produced by preheating was studied.
}
The linear equation (\ref{eq:chiEoM}) can be decomposed into Fourier modes by expanding $\chi$ in terms of the creation and annihilation operators as
\begin{equation}
    \chi (t, \vec{x}) = \int \frac{d^3 k}{(2 \pi)^3} \left( a_{\vec{k}} \chi_k (t) + a^{\dagger}_{-\vec{k}} \chi^*_k (t) \right) e^{i \vec{k} \cdot \vec{x}} ,
    \label{eq:chiquantiz}
\end{equation}
where the creation and annihilation operators satisfy the following commutation relations:
\begin{align}
    [a_{\vec{k}}, a_{\vec{k}'}^\dagger] = ( 2\pi )^3 \delta^{(3)} (\vec{k} - \vec{k}') , \quad
    [a_{\vec{k}}, a_{\vec{k}'}] = [a_{\vec{k}}^\dagger, a_{\vec{k}'}^\dagger] = 0 ,
\end{align}
and each mode satisfies
\begin{equation}
    \ddot{\chi}_k + 3H \dot{\chi}_k + \omega_k^2 \chi_k = 0 ,
    \label{eq:chikEoM}
\end{equation}
where $\omega_k = \sqrt{(k/a)^2 + g^2 \phi^2}$.
The equation of motion includes the dissipation term, so $\chi$ and $\dot{\chi}$ are not a canonical pair. To make it more transparent,
we switch to conformal time $\tau$ defined by $dt = a d\tau$, and rescale the field as $\tilde{\chi} \equiv a\chi$.
Defining $\tilde{\chi}_k \equiv a\chi_k$, the equation of motion (\ref{eq:chikEoM}) then becomes
\begin{align}
    \tilde{\chi}''_k + \tilde{\omega}_k^2 \tilde{\chi}_k = 0 ,
    \label{eq:chitildeEoM}
\end{align}
where $\tilde{\omega}_k = \sqrt{k^2 + a^2 g^2 \phi^2 - a''/a}$, and $'$ means conformal time derivative.
Now, the variable $\tilde{\chi}'$ is the canonical momentum of $\tilde{\chi}$.
Eq.~(\ref{eq:chiquantiz}) becomes
\begin{align}
    \tilde{\chi} (\tau, \vec{x}) = \int \frac{d^3 k}{(2 \pi)^3} \left( a_{\vec{k}} \tilde{\chi}_k (\tau) + a^{\dagger}_{-\vec{k}} \tilde{\chi}^*_k (\tau) \right) e^{i \vec{k} \cdot \vec{x}}.
\end{align}
We impose the canonical commutation relation $[\tilde{\chi} (\tau , \vec{x}) , \tilde{\chi}' (\tau , \vec{y})] = i \delta^{(3)} (\vec{x} - \vec{y})$, which implies the following normalization condition:
\begin{align}
    \tilde{\chi}_k \tilde{\chi}'^*_k - \tilde{\chi}^*_k \tilde{\chi}'_k = i .
\end{align}
The solution of Eq.~(\ref{eq:chitildeEoM}) can be written in the adiabatic basis $\exp(\pm i \int \tilde{\omega}_k d\tau)$ as
\begin{align}
    \tilde{\chi}_k (\tau)
    = \frac{\alpha_k (\tau)}{\sqrt{2\tilde{\omega}_k(\tau)}} \exp \left( - i \int_{\tau_{\rm ini}}^{\tau} \tilde{\omega}_k (\tau') d\tau' \right)
    + \frac{\beta_k (\tau)}{\sqrt{2\tilde{\omega}_k(\tau)}} \exp \left( + i \int_{\tau_{\rm ini}}^{\tau} \tilde{\omega}_k (\tau') d\tau' \right)
    ,
\end{align}
where $\tau_{\rm ini}$ denotes the initial time, and $\alpha_k$ and $\beta_k$ are  time-dependent Bogoliubov coefficients that satisfy the normalization condition, $|\alpha_k|^2 - |\beta_k|^2 = 1$, and evolve according to
\begin{align}
    \frac{d \alpha_k}{d \tau} &= \frac{\beta_k}{2 \tilde{\omega}_k}  \frac{d \tilde{\omega}_k}{d \tau} \exp \left( + 2 i \int_{\tau_{\rm ini}}^{\tau} \tilde{\omega}_k (\tau') d\tau' \right)
    ,\label{eq:BogolEoM1}
    \\
    \frac{d \beta_k}{d \tau} &= \frac{\alpha_k}{2 \tilde{\omega}_k}  \frac{d \tilde{\omega}_k}{d \tau} \exp \left( - 2 i \int_{\tau_{\rm ini}}^{\tau} \tilde{\omega}_k (\tau') d\tau' \right)
    .
    \label{eq:BogolEoM}
\end{align}
We set the initial conditions $\alpha_k (\tau_{\rm ini}) = 1$ and $\beta_k (\tau_{\rm ini}) = 0$, and we define the vacuum state $| 0 \rangle$ by $a_{\vec{k}} |0 \rangle = 0$. The initial $\tau_{\rm ini}$ is taken well before the first inflaton zero-crossing.
Practically, the computational complexity can be reduced by absorbing the fast phase oscillations via  the variable transformation, 
$A_k = \alpha_k \exp \left( - i \int \tilde{\omega}_k d\tau \right)$ and $B_k = \beta_k \exp \left( + i \int \tilde{\omega}_k d\tau \right)$.
Eqs.~(\ref{eq:BogolEoM1}) and (\ref{eq:BogolEoM}) are then equivalent to
\begin{align}
    \frac{d A_k}{d \tau} &= \frac{B_k}{2 \tilde{\omega}_k}  \frac{d \tilde{\omega}_k}{d \tau} -i \tilde{\omega}_k A_k
    ,\\
    \frac{d B_k}{d \tau} &= \frac{A_k}{2 \tilde{\omega}_k}  \frac{d \tilde{\omega}_k}{d \tau} + i \tilde{\omega}_k B_k
    .
    \label{eq:BogolEoMmod}
\end{align}

We define the physical number density of $\chi$ as
\begin{align}
    n_{\chi}(\tau) &\equiv a^{-3} (\tau) \int \frac{d^3 \vec{k}}{(2\pi)^3} n_{\chi ,k}(\tau) ,\\
    n_{\chi ,k}(\tau) &\equiv \frac{1}{2\tilde{\omega}_k} \left( |\tilde{\chi}'_k (\tau)|^2 + \tilde{\omega}_k^2 |\tilde{\chi}_k (\tau)|^2 \right) - \frac{1}{2} = |\beta_k (\tau)|^2 = |B_k (\tau)|^2 ,
\end{align}
where $n_{\chi ,k}$ is the comoving number spectrum.
The term $-\frac{1}{2}$ subtracts the  vacuum contribution.
Note that $|\beta_k (\tau)|^2$ corresponds to the expectation value of the number operator $\langle 0| b_{\vec{k}}^\dagger (\tau) b_{\vec{k}} (\tau) |0\rangle$ after the renormalization where $b_{\vec{k}}^{(\dagger)}$ is the annihilation(creation) operator for the instantaneous positive frequency mode, $b_{\vec{k}} = \alpha_k (\tau) a_{\vec{k}} + \beta_k (\tau) a_{-\vec{k}}^\dagger$.
As is clear from Eq.~(\ref{eq:BogolEoMmod}), $n_{\chi ,k}$ is approximately conserved when the evolution is adiabatic, $|\dot{\omega}_k / \omega_k^2| \ll 1$, and it changes only during non-adiabatic intervals, typically around the inflaton zero-crossing.

The produced number density of $\chi$ particles after the first passage of the inflaton $\phi=0$ is~\cite{Kofman:1997yn}
\begin{equation}
    n_\chi \simeq \left(\frac{p_*}{2\pi}\right)^3,
\end{equation}
where $p_*$ is the typical momentum of the produced particle given by
\begin{equation}
    p_* \equiv \sqrt{g \phi_* m_\phi} = m_\phi q^{1/4},~~~~~~q \equiv \frac{g^2 \phi_*^2}{m_\phi^2}.
\end{equation}
Here $\phi_*$ is the inflaton oscillation amplitude, and  $\phi_* \sim M_{\rm pl}$ for chaotic inflation.
Unless produced $\chi$ particles promptly decay into  lighter species,  parametric resonance enhances the $\chi$ production. 
However, we are mainly interested in relatively short duration after inflation, i.e., period during which the inflaton oscillates a few times. In such a short period, the parametric resonance effect is not significant.

\begin{figure}[!t]
\begin{center}
    \includegraphics[width=0.49\textwidth]{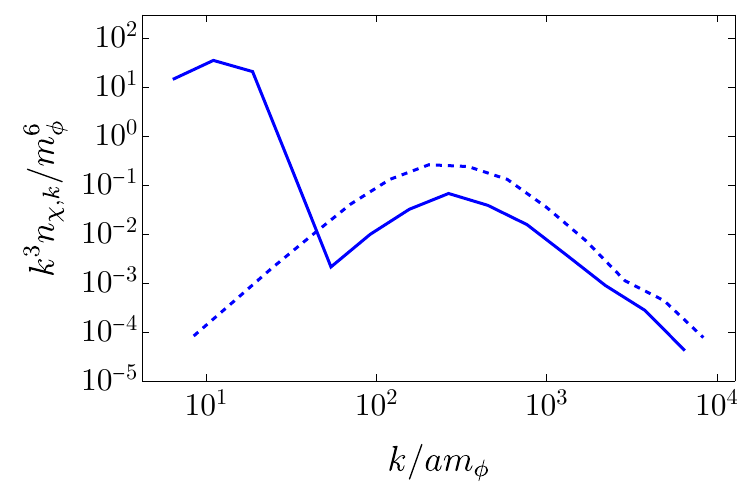}
    \includegraphics[width=0.49\textwidth]{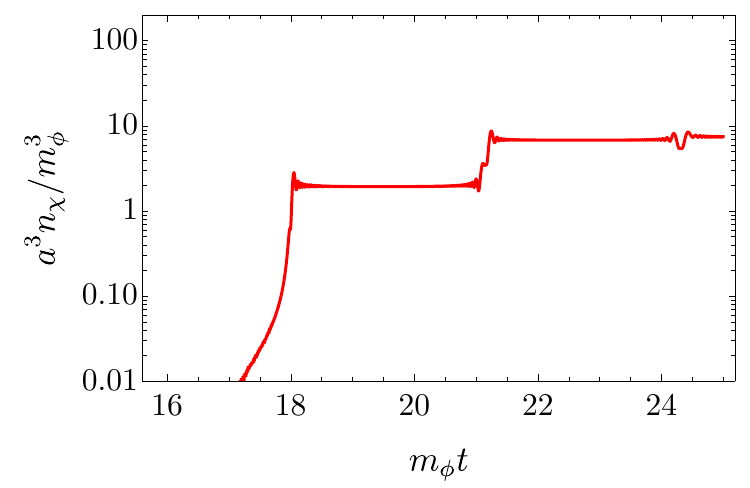}
\end{center}
\caption{%
    The left panel shows
    the power spectrum of the comoving number spectrum, $n_{\chi ,k}$,
    before (dashed line) and after (solid line) the first inflaton zero-crossing.
    The momentum in the horizontal axis is normalized by the scale factor with $a(\tau_{\rm ini}) = 1$.
    The right panel shows
    the time evolution of the comoving number density $a^3 n_\chi$. 
    The inflaton mass and the coupling constant are set to be $m_\phi = 10^{-6} M_{\rm pl}$ and $g=10^{-3}$, respectively.
}
\label{fig: number_spectrum}
\end{figure}

Fig.~\ref{fig: number_spectrum} shows the numerical results of the time evolution of the power spectrum of the number density, $n_{\chi ,k}$ (left panel), and the number density, $n_\chi$ (right panel).
Here we set $m_\phi = 10^{-6}\, M_{\rm pl}$, $g = 10^{-3}$ and choose $\tau_{\rm ini}$ to be when $\phi = \sqrt{2}\, M_{\rm pl}$.
The IR cutoff of the comoving momentum is taken well above $aH$, and the UV cutoff is well above the effective-mass scale, $a g \phi_*$.
The characteristic momentum $p_*$ is included in this range, i.e.,
\begin{align}
    H < \frac{k_{\rm IR}}{a} < p_* < g \phi_* < \frac{k_{\rm UV}}{a} .
\end{align}
The scale factor here is rescaled so that $a(\tau_{\rm ini}) = 1$.
The value of $q$ around the first zero crossing of the inflaton is $q \simeq 5 \times 10^4$.
We can see that, after the first passage, the spectrum peaks at $k/a \sim 10\, m_\phi$ which is near the typical momentum estimated, $p_* \simeq 15\, m_\phi$. 
Also, the $\chi$ particles are efficiently produced when the inflaton passes the origin.

%%%%%%%%%%%%%%%%%%%%%%%%%%%%%%%%%%%%%%
%%%%%%%%%%%%%%%%%%%%%%%%%%%%%%%%%%%%%%

%%%%%%%%%%%%%%%%%%%%%%%%%%%%%%%%%%%%%%
\subsection{Numerical results of Higgs dynamics}
\label{subsec:NumNoneq}
%%%%%%%%%%%%%%%%%%%%%%%%%%%%%%%%%%%%%%

The $\chi$ particles produced during preheating affect the Higgs dynamics through the interaction in Eq.~(\ref{eq:chisigma}).
The contribution to the Higgs effective potential induced by the $\chi$ particles can be approximated as
\begin{equation}
    V_{\chi\sigma} = - \frac{1}{2} \lambda_{\chi\sigma} \langle \chi^2 (\tau) \rangle \sigma^2,
\end{equation}
since $\chi$ becomes highly occupied and then $\langle \chi^2 \rangle$ acts like a classical background.
The expectation value $\langle \chi^2 \rangle$ is calculated as
\begin{align}
    \langle \chi^2 (\tau) \rangle
    &= a^{-2} (\tau) \langle 0 | (\tilde{\chi}^2 (\tau) - \tilde{\chi}^2 (\tau_{\rm ini})) | 0 \rangle \\
    &= a^{-2} (\tau) \int \frac{d^3 \vec{k}}{(2\pi)^3} \frac{1}{\tilde{\omega}_k (\tau)} \left( |\beta_k (\tau)|^2 + {\rm Re}\,{\cal A} (\tau) \right) 
\end{align}
where the initial expectation value is subtracted for the renormalization and ${\cal A}$ represents a fast oscillating part,
\begin{align}
    {\cal A} (\tau) \equiv A_k(\tau) B_k^*(\tau) =  \alpha_k \beta_k^* \exp \left( - 2 i \int_{\tau_{\rm ini}}^{\tau} \tilde{\omega}_k (\tau') d\tau' \right).
\end{align}
The oscillatory term appears from the interference between the produced pairs of the positive-frequency modes and negative-frequency modes because the background inflaton deforms the vacuum of $\chi$ into one full of the correlated particle pairs, the squeezed state.
In a general case of high-scale inflation, such an oscillatory term has a high frequency of the same order as the inflaton amplitude up to the coupling constant and can be dropped by temporal averaging over the time scale of the $\sigma$ dynamics ($g\phi \gg \sqrt{c} H$), while the amplitude of the oscillation is ${\cal O}(1)$~\cite{Kofman:1997yn}.
As we show later, dropping ${\rm Re}\, {\cal A}$ part does not change the $\sigma$ dynamics significantly.
For later use, we define
\begin{align}
    \overline{\langle \chi^2 \rangle}
    \equiv
    a^{-2} \int \frac{d^3 \vec{k}}{(2\pi)^3} \frac{n_{\chi , k}}{\tilde{\omega}_k}
    .
    \label{eq:def_chisq_bar}
\end{align}

\begin{figure}[!t]
\begin{center}
\includegraphics[width=0.6\textwidth]{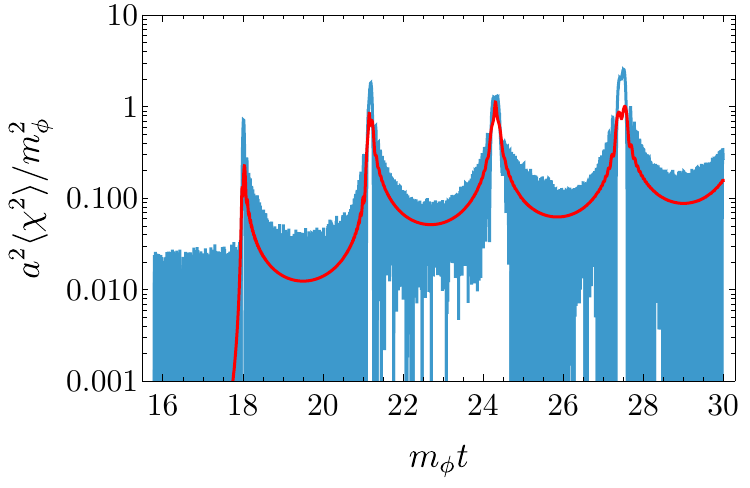}
\end{center}
\caption{%
    The numerical result of the time evolution of $\langle \chi^2 \rangle$ corresponding to the result in Fig.~\ref{fig: number_spectrum} (blue line)
    and
    the result of $\overline{\langle \chi^2 \rangle}$ defined in Eq.~(\ref{eq:def_chisq_bar}), in which the oscillatory term ${\rm Re}\, {\mathcal A}$ is subtracted (red line).
}
\label{fig: chisq}
\end{figure}

Fig.~\ref{fig: chisq} shows the numerical result of the time evolution of $\langle \chi^2 \rangle$.
The result in the left panel intensely oscillates due to the ${\rm Re}\,{\cal A}$ term, which is not seen in $\overline{\langle \chi^2 \rangle}$ shown in the right panel.

After the negative coupling term dominates over the negative Hubble mass term, the potential minimum remains far from the origin.
If this happens during the first oscillation of the inflaton, $\sigma$ is expected to be trapped around the potential minimum.
The condition that the negative coupling term dominates over the Hubble mass term is written as
\begin{equation}
    \lambda_{\chi\sigma}  \overline{\left< \chi^2\right>} \gtrsim c H^2,
    \label{eq:trap_cond}
\end{equation}
after $\chi$ is amplified by the first zero-crossing of the inflaton.
Since the typical mode $p_*$ is specifically amplified during the preheating process, $\overline{\left< \chi^2 \right>}$ is estimated as
\begin{equation}
    \overline{\left< \chi^2 \right>} \sim \frac{n_\chi}{\omega_{k_*}} ,
    \label{chi2}
\end{equation}
where $k_* / a = p_*$.
Although the value varies with $|\phi|$, there is a typical lower limit,
\begin{equation}
    \overline{\left< \chi^2 \right>} \gtrsim \frac{m_\phi^2}{8\pi^3} q^{1/4} .
\end{equation}
The minimum value is sufficient to derive a conservative condition for keeping $\sigma$ away from the origin.
Noting that $H \sim 0.1 m_\phi$ at the first passage of $\phi=0$, the condition (\ref{eq:trap_cond}) is satisfied when
\begin{equation}
    \lambda_{\chi\sigma} \gtrsim c q^{-1/4}.
    \label{eq:coupling_condition}
\end{equation}

\begin{figure}[!t]
\begin{center}
    \includegraphics[width=0.49\textwidth]{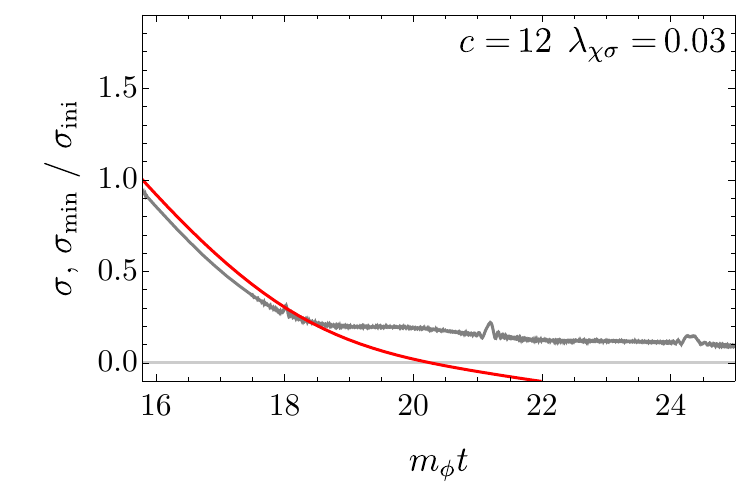}
    \includegraphics[width=0.49\textwidth]{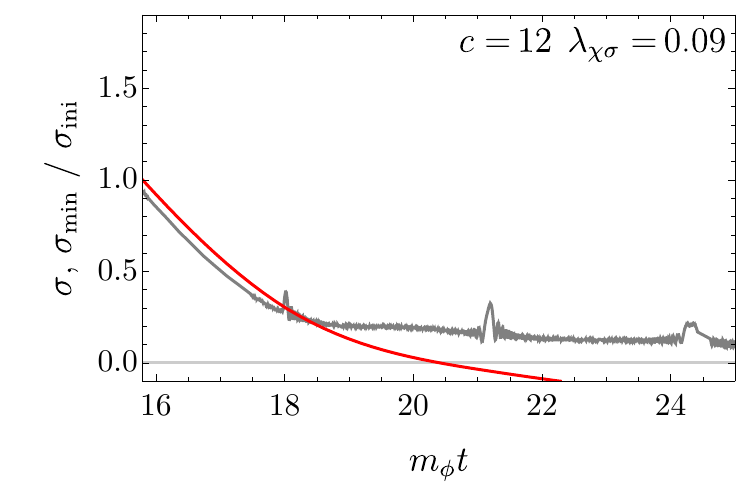}
    \includegraphics[width=0.49\textwidth]{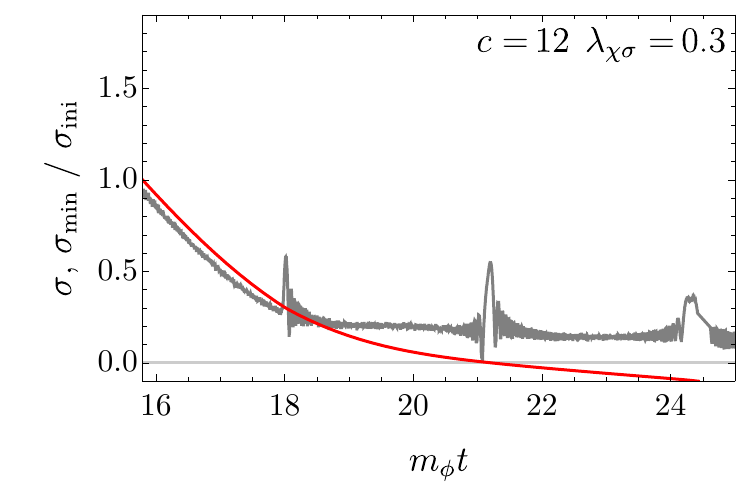}
    \includegraphics[width=0.49\textwidth]{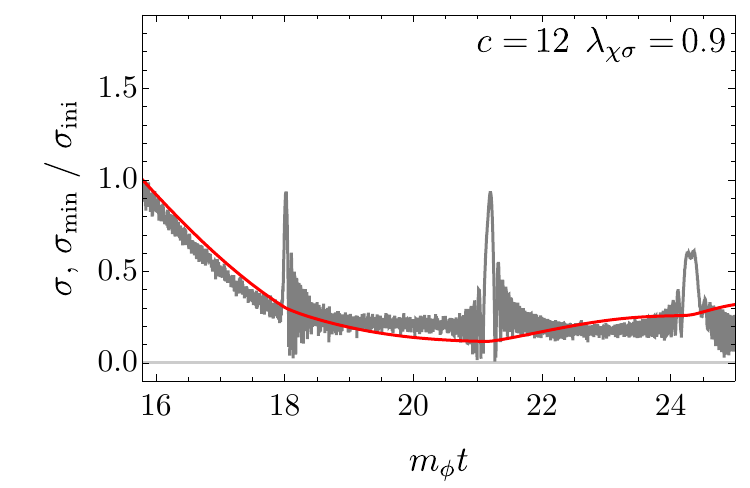}
\end{center}
\caption{%
    The time evolution of $\sigma$ (red) and the potential minimum $\sigma_{\rm min}$ (gray) normalized by the initial value $\sigma_{\rm ini} = \sigma |_{t=t_{\rm ini}}$ for $c = 12$.
    The coupling constant $\lambda_4$ is set to an appropriate value for each panel.
}
\label{fig: sigma_c12}
\end{figure}

\begin{figure}[!t]
\begin{center}
    \includegraphics[width=0.49\textwidth]{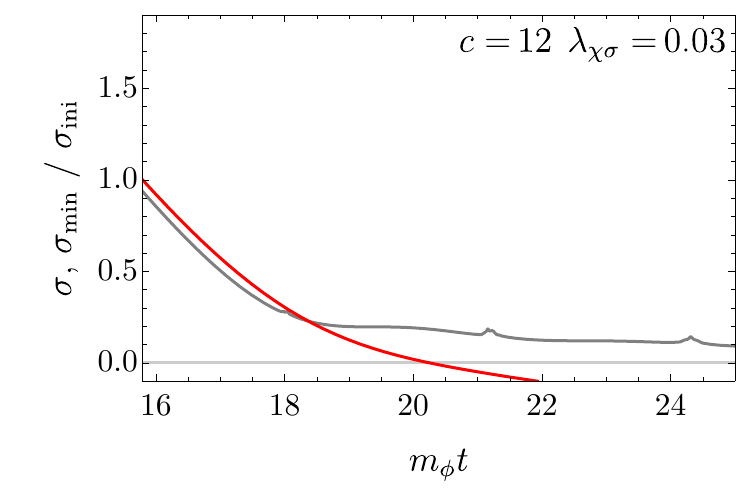}
    \includegraphics[width=0.49\textwidth]{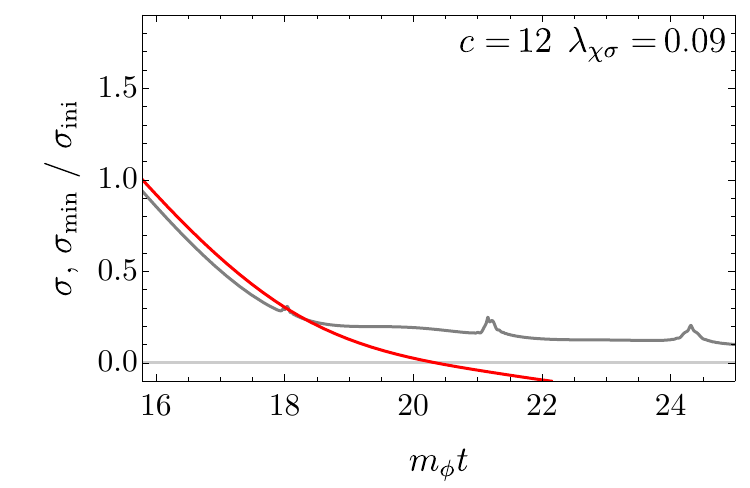}
    \includegraphics[width=0.49\textwidth]{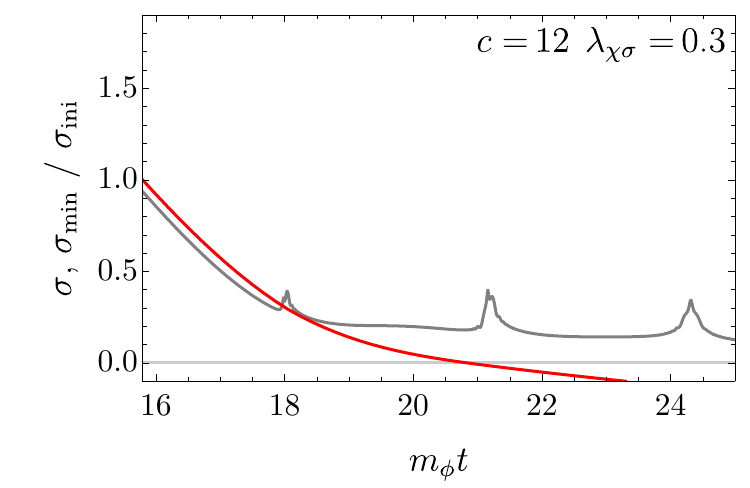}
    \includegraphics[width=0.49\textwidth]{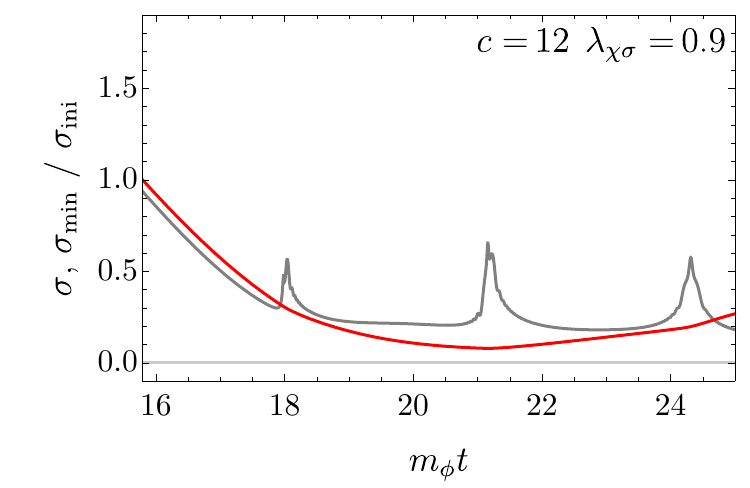}
\end{center}
\caption{%
    Same as Fig.~\ref{fig: sigma_c12} except for subtracting ${\rm Re} {\cal A}$ contribution from $\langle \chi^2 \rangle$.
}
\label{fig: sigmawoReA_c12}
\end{figure}

\begin{figure}[!t]
\begin{center}
    \includegraphics[width=0.49\textwidth]{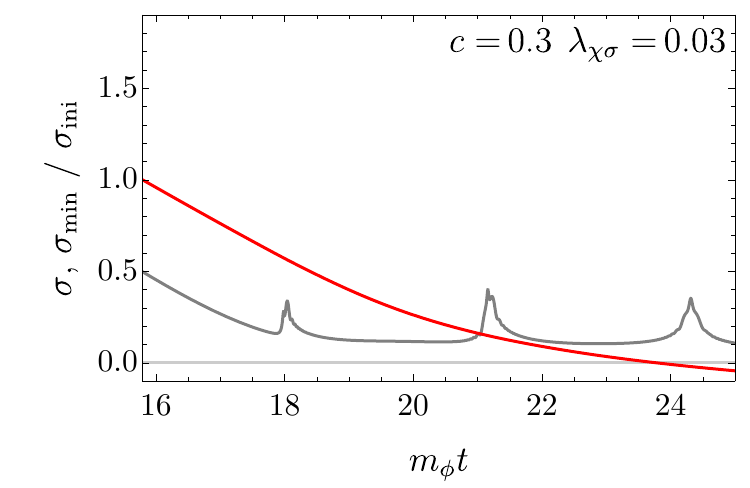}
    \includegraphics[width=0.49\textwidth]{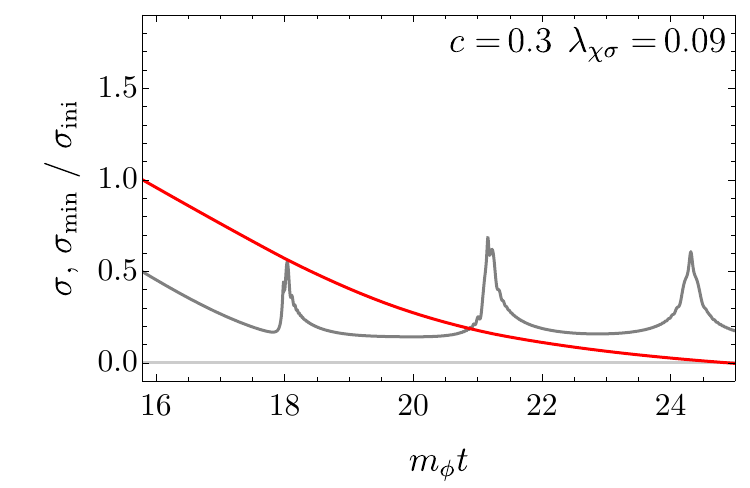}
    \includegraphics[width=0.49\textwidth]{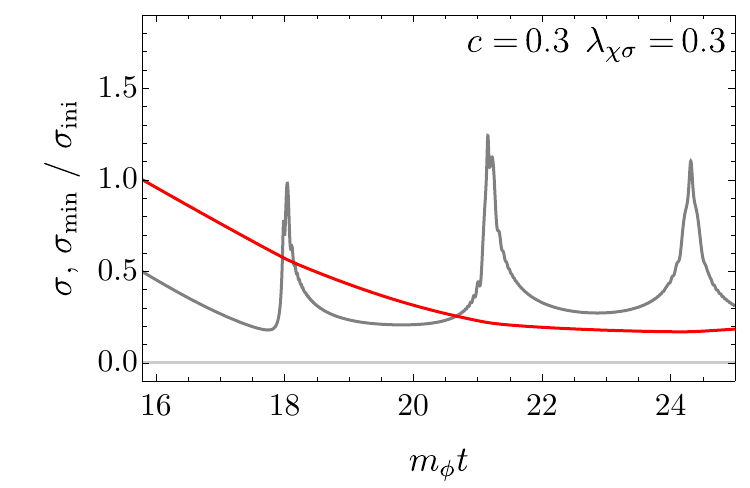}
    \includegraphics[width=0.49\textwidth]{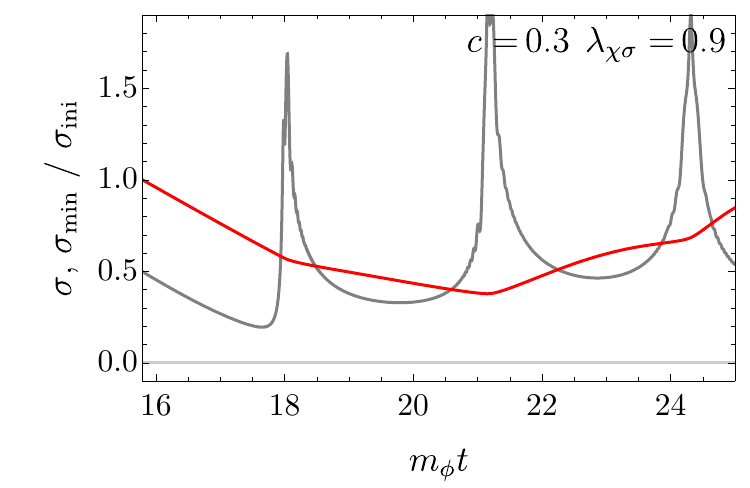}
\end{center}
\caption{%
    Same as Fig.~\ref{fig: sigmawoReA_c12} except for $c = 0.3$.
}
\label{fig: sigmawoReA_c03}
\end{figure}

We calculated the evolution of the Higgs by solving
\begin{equation}
    \ddot{\sigma} + 3H\dot{\sigma}
    + \lambda_4 \sigma^3 - \left[ c H^2 + \lambda_{\chi\sigma} \langle \chi^2 (t) \rangle \right] \sigma
    = 0
    .
    \label{sigma_eom_preheating}
\end{equation}
For the initial conditions $\sigma_{\rm ini}$ and $\dot{\sigma}_{\rm ini}$, we reuse the $\sigma (t_{\rm ini})$ and $\dot{\sigma} (t_{\rm ini})$ results
obtained in Sec.~\ref{subsec: Inflation}
for the corresponding value of $c$.
The initial time $t_{\rm ini}$ is set such that $\phi = \sqrt{2}\, M_{\rm pl}$ at $t = t_{\rm ini}$ and so the calculation starts well before the onset of inflaton oscillations.
The coefficient $\lambda_4$ is chosen to satisfy $\lambda_{\chi\sigma}^2 < \lambda_4 < \lambda_{\chi\sigma}$ so that the potential is bounded from below and the fluctuation of $\sigma$ is not large enough to restore the symmetry, as we discuss later, although the results do not depend on the choice of $\lambda_4$ when solving Eq.~(\ref{sigma_eom_preheating}).
Fig.~\ref{fig: sigma_c12} shows the numerical results for $c=12$.
Each of the four panels shows the result for a different value of $\lambda_{\chi\sigma}$, and the red (gray) lines show the time evolution of $\sigma$ ($\sigma_{\rm min}$) normalized by the initial field value $\sigma_{\rm ini}$.
We can see that $\sigma$ traces the potential minimum in the case of $\lambda_{\chi\sigma} = 0.9$ but not in the other cases.
The result may be understood from the condition~\eqref{eq:coupling_condition}: since $q \simeq 5 \times 10^4$ and $c = 12$, the condition for symmetry non-restoration is $\lambda_{\chi\sigma} \gtrsim {\cal O}(1)$.
We show the result when we replace $\langle \chi^2 \rangle$ by $\overline{\langle \chi^2 \rangle}$ in Fig.~\ref{fig: sigmawoReA_c12}.
Here, we have confirmed that the subtraction of the oscillating part, $\mathrm{Re} \, \mathcal{A}$, from $\langle \chi^2 \rangle$ hardly changes the dynamics of $\sigma$.
Fig.~\ref{fig: sigmawoReA_c03} shows results with $c = 0.3$.
Here, $\sigma$ does not reach the symmetry restoration point during a few inflaton oscillations in the cases with $\lambda_{\chi\sigma} = 0.3$ and $0.9$.

So far we have neglected the backreaction of the Higgs $\sigma$ on the $\chi$ dynamics. Let us comment on the condition for which the above calculations are valid.
First, the Higgs vacuum expectation value contributes to the mass of $\chi$ through the negative portal coupling.
Then the negligible backreaction requires
\begin{align}
    p_\star^2 \gg \lambda_{\chi\sigma}\langle \sigma^2 \rangle ,
\end{align}
during the period of our interest.
Assuming that the mean field $\langle \sigma \rangle$ traces $\sqrt{(\lambda_{\chi\sigma} / \lambda_4) \overline{\langle \chi^2 \rangle}}$ and that fluctuations around the mean field, $\delta \sigma$, is small, this condition becomes
\begin{align}
    p_\star^2 \gg \frac{\lambda_{\chi\sigma}^2}{\lambda_4} \overline{\langle \chi^2 \rangle} .
\end{align}
Using $\overline{\langle \chi^2 \rangle} \sim p_\star^2 / 8\pi^3$, we find that $\lambda_4$ should satisfy
\begin{align}
    \lambda_4 \gg \frac{\lambda_{\chi\sigma}^2}{8\pi^3} ,
\end{align}
which is naturally realized as we require the Hamiltonian to be bounded from below, $\lambda_4 \lambda_\chi > \lambda_{\chi\sigma}^2$.
This condition also ensures that the effective potential of $\chi$ is minimized at $\chi=0$.
Second, the Higgs fluctuation $\delta\sigma$ may be efficiently produced  through scatterings of $\chi$ particles, which might potentially affect the whole dynamics.
If the fluctuation $\delta \sigma$ becomes large enough, it would restore the symmetry of $\sigma$.
To avoid the symmetry restoration of $\sigma$, we require $\left<\delta\sigma^2\right> < \sigma^2 \sim (\lambda_{\chi\sigma}/\lambda_4)\overline{\left<\chi^2\right>}$.
The maximally efficient production case is $\rho_{\delta\sigma}\sim\rho_\chi$ or $\left<\delta\sigma^2\right> \sim \overline{\left<\chi^2\right>}$, thus $\lambda_{\chi\sigma} > \lambda_4$ is a sufficient condition to avoid the symmetry restoration.
Let us more explicitly evaluate it.
The fluctuation $\delta \sigma$ is produced through the interaction with $\chi$, and the cross section of the process $\chi \chi \rightarrow \sigma \sigma$ is estimated as $\sigma_{\chi\chi \rightarrow \sigma\sigma} \simeq \lambda_{\chi\sigma}^2 / 32\pi p_\star^2$.
The rate of the process is then given by
\begin{align}
    \Gamma_{\chi\chi \rightarrow \sigma\sigma} = n_\chi \sigma_{\chi\chi \rightarrow \sigma\sigma} \sim \frac{\lambda_{\chi\sigma}^2}{256 \pi^4} p_\star \sim \frac{\lambda_{\chi\sigma}^2}{256 \pi^4} q^{1/4} m_\phi .
\end{align}
Comparing this with the Hubble parameter at the first zero-crossing of the inflaton, $H \sim 0.1 m_\phi$, we find that $\Gamma_{\chi\chi \rightarrow \sigma\sigma} < H$ for $\lambda_{\chi\sigma} \lesssim 10$.
Thus, the fluctuation of $\sigma$ is not efficiently generated through the interaction with $\chi$ during the period of our interest, while $\lambda_{\chi\sigma} > \lambda_4$ is still required to keep $\delta \sigma$ small after the thermalization of $\chi$ as discussed in Sec.~\ref{sec:ther}.

%%%%%%%%%%%%%%%%%%%%%%%%%%%%%%%%%%%%%%
%%%%%%%%%%%%%%%%%%%%%%%%%%%%%%%%%%%%%%

%%%%%%%%%%%%%%%%%%%%%%%%%%%%%%%%%%%%%%
%%%%%%%%%%%%%%%%%%%%%%%%%%%%%%%%%%%%%%

%%%%%%%%%%%%%%%%%%%%%%%%%%%%%%%%%%%%%%
%%%%%%%%%%%%%%%%%%%%%%%%%%%%%%%%%%%%%%

%%%%%%%%%%%%%%%%%%%%%%%%%%%%%%%%%%%%%%
%%%%%%%%%%%%%%%%%%%%%%%%%%%%%%%%%%%%%%

%%%%%%%%%%%%%%%%%%%%%%%%%%%%%%%%%%%%%%
\section{Implications for PQ scalars in Axion Models}
\label{sec:PQ}
%%%%%%%%%%%%%%%%%%%%%%%%%%%%%%%%%%%%%%
So far, we have examined how the radial component of a generic complex scalar field, $\sigma$, evolves under a negative Hubble-induced mass term, a negative thermal mass, and an effective mass generated by particles produced non-thermally during preheating, and we have discussed whether the associated symmetry is restored. We now turn to the implications of these results when the complex scalar is identified with the PQ field.

As emphasized in the Introduction, the QCD axion generally suffers from the domain wall problem. Whether the U(1)$_{\rm PQ}$ symmetry is restored after inflation is therefore directly linked to whether the PQ mechanism for solving the strong CP problem leads to cosmological difficulties. Our results indicate that if another scalar field coupled to the PQ scalar is produced during preheating, the PQ field can be kept away from the origin throughout the post-inflationary epoch. In such a case, the production of topological defects such as strings and domain walls can be avoided. We note that the Standard Model Higgs boson can play a role of such a coupled scalar field, and hence we do not need to extend a model for this mechanism to work.

Another important cosmological issue for the QCD axion is the isocurvature constraint on the inflationary scale. In the pre-inflationary scenario, this bound severely disfavors high-scale inflation~\cite{Planck:2018jri}. However, our mechanism provides a realization of the idea originally proposed in Refs.~\cite{Linde:1990yj,Linde:1991km}: the PQ scalar is driven to a large field value during inflation and subsequently relaxes to its present value, thereby suppressing axionic isocurvature fluctuations.\footnote{
Note that axion fluctuations are enhanced at small scales as the PQ field value gradually decreases~\cite{Kobayashi:2016qld}.
Also, PQ breaking terms suppressed by the Planck mass may make the axion sufficiently heavy to suppress  isocurvature perturbations~\cite{Higaki:2014ooa}.
} This requires the PQ quartic coupling to be extremely small in non-supersymmetric models~\cite{Linde:1991km,Kasuya:1996ns}. 
More generally, the dynamics of the PQ scalar depend on the detailed structure of the PQ sector, including how the field is stabilized, how many PQ scalars are present, and whether the underlying theory is supersymmetric. 
A potential problem of these scenarios was dynamical symmetry restoration after inflation and the resulting domain wall formation due to the oscillation of the PQ scalar, but our mechanism provides a way to avoid it.
It is also worth noting that a passage of the field through the origin does not automatically imply symmetry restoration or string formation~\cite{Graham:2025iwx}.

In the post-inflationary scenario, the axion is usually associated with a string-domain wall network. In alternative scenarios, however, one may instead consider axion domain walls without attached strings. Such stringless domain walls provide a viable candidate, for example, to explain isotropic cosmic birefringence~\cite{Takahashi:2020tqv,Kitajima:2022jzz,Gonzalez:2022mcx}. This situation arises naturally if the PQ quartic coupling is of order unity: the field displacement during inflation is of order the Hubble scale, while axion fluctuations are also of order the Hubble parameter, so the phase fluctuations are of order unity. This naturally leads to the formation of stringless axion domain walls.%
\footnote{
    The domain wall network is stable against a population bias if the initial fluctuations are of the inflationary origin~\cite{Gonzalez:2022mcx}.
} 
Our mechanism can be applied to realize this type of dynamics.

%%%%%%%%%%%%%%%%%%%%%%%%%%%%%%%%%%%%%%
\section{Conclusions}
\label{sec:conc}
%%%%%%%%%%%%%%%%%%%%%%%%%%%%%%%%%%%%%%

We have studied the post-inflationary evolution of a Higgs field $S$ that acquires a large expectation value during inflation due to a negative Hubble-induced mass term, and shown that symmetry restoration after inflation is not inevitable. There are well-motivated scenarios in which $S$ never reaches the origin and topological defects are not regenerated.

For example, when the post-inflationary minimum is determined by the balance between a negative Hubble-induced mass term and a quartic coupling term, the oscillation amplitude grows relative to the distance from the origin, and the field inevitably crosses it. If instead the minimum is stabilized by higher-dimensional operators, the field can adiabatically follow the evolving minimum without crossing the origin, provided the operator dimension is sufficiently high and depending on the background equation of state.

An effective negative thermal mass is also known to prevent a zero-crossing even with a quartic potential, provided that the light species coupled to the Higgs sector thermalizes instantaneously. However, whether complete thermalization is achieved immediately after inflation is not obvious, which motivates a more realistic treatment of the early non-equilibrium stage.

We therefore analyzed the effect of preheating, which naturally generates a non-thermal distribution of the coupled light field before thermalization. The resulting finite-density contribution acts as an effective negative mass for the Higgs field and keeps the field in the broken phase, preventing a zero crossing. Consequently, even in renormalizable models where, at low energies, the potential is determined by the quadratic mass term and the quartic self-interaction, non-thermally produced scalars during preheating can maintain the broken phase and prevent the formation of topological defects.

Our analyses are quite general and have broad applications to concrete models in which topological defects can be problematic.
While the case of PQ symmetry has been discussed in the previous section, other examples are grand unified theories~\cite{Langacker:1980js}, models with discrete flavor symmetries for neutrino masses~\cite{Altarelli:2010gt,King:2014nza}, spontaneous CP violation where CP domain walls necessarily exist~\cite{Nelson:1983zb,Barr:1984qx} (see Ref.~\cite{Asadi:2022vys} for a recent discussion and Refs.~\cite{Murai:2024alz,Murai:2024bjy} for solutions to the domain wall problem), and extended Higgs sectors such as the $Z_2$-symmetric two-Higgs-doublet model~\cite{Branco:2011iw} and the next-to-minimal supersymmetric Standard Model (NMSSM)~\cite{Ellwanger:2009dp}.
It is often assumed that topological defects are inflated away or that the discrete symmetry is explicitly broken by a small amount to make domain walls unstable.
However, the former solution constrains the relation between the inflation scale and the symmetry-breaking scale, and the latter solution often makes models complicated.
Our findings imply that dangerous topological defect formation can be avoided by carefully considering the Higgs dynamics after inflation, without introducing additional model ingredients.
This opens up the possibility that models previously thought to be impossible could be revived.

%%%%%%%%%%%%%%%% Note %%%%%%%%%%%%%%%%%
\section*{Note Added}
During the early stages of this work, we became aware that another group was independently studying a related subject~\cite{Nakagawa:2025suc}. After discussions with them, we agreed to coordinate and submit our papers on the same day. We would like to thank them for the helpful interaction.
%%%%%%%%%%%%%%%%%%%%%%%%%%%%%%%%%%%%%%

%%%%%%%%%%%%%%%%%%%%%%
\section*{Acknowledgment}
%%%%%%%%%%%%%%%%%%%%%%

This work was supported by JSPS KAKENHI (Grant Numbers 24K07010 [KN], 25H02165 [FT], 25KJ0564 [JL], 23KJ0088 [KM], and 24K17039 [KM]).
This work was also supported by World Premier International Research Center Initiative (WPI), MEXT, Japan.
One of us (JL) is supported by Graduate Program on Physics for the Universe (GP-PU), Tohoku University.

\appendix

%%%%%%%%
\section{Conservation of comoving number density of the Higgs}
\label{app:cons}
In this Appendix, we show that the comoving number density of the Higgs field is conserved and prove Eq.~(\ref{eq:delta sigma}). Let us suppose that the Higgs field oscillates around $\sigma_m(t)$ and write $\sigma(t) = \sigma_m(t) + \delta\sigma(t)$. Note that $\sigma_m(t)$ is slightly different from $\sigma_{\rm min}(t)$~\cite{Dine:1995kz}. For example, in the model of Sec.~\ref{sec:Hub}, it is given by
\begin{align}
    \sigma_m(t) = \left[1 + \frac{6p(n-2)-2n}{cp^2(n-2)^2}\right]^{\frac{1}{n-2}}\sigma_{\rm min}(t),
\end{align}
so that it satisfies the equation of motion $\ddot\sigma_m+3H\dot\sigma_m+V'(\sigma_m)=0$, where we assumed $H=p/t$. 
Then $\delta\sigma$ satisfies
\begin{align}
    \ddot{\delta\sigma} + 3H\dot{\delta\sigma} + m_\sigma^2(t) \delta\sigma = 0,
    \label{eq:deltasigma}
\end{align}
where $m_\sigma(t)$ is the time-dependent mass around $\sigma=\sigma_m(t)$. In the model of Sec.~\ref{sec:Hub} it is given by $m_\sigma(t) \sim \sqrt{c}\,H$, while in the model of Sec.~\ref{sec:ther} it is $m_\sigma(t) \sim\sqrt{\lambda_{\chi\sigma}}\,T$, but here we keep it in a general form. 
Our assumption is that $m_\sigma \gg H$
and $|\dot m_\sigma/m_\sigma| \ll m_\sigma$.
By taking oscillation average, we soon find $\left<\dot{\delta\sigma}^2\right> \simeq m_\sigma^2 \left<\delta\sigma^2\right>$.
Thus the energy density is given by $\rho_\sigma = \left(\dot{\delta\sigma}^2 + m_\sigma^2\delta\sigma^2\right)/2 \simeq m_\sigma^2\left<\delta\sigma^2\right>$. 
By using this relation, Eq.~(\ref{eq:deltasigma}) yields
\begin{align}
    \dot\rho_\sigma = -3H\dot{\delta\sigma}^2 + \dot m_\sigma m_\sigma \delta\sigma^2
    \simeq \left(-3H + \frac{\dot m_\sigma}{m_\sigma}\right)\rho_\sigma.
\end{align}
This is rewritten as
\begin{align}
    \frac{d}{dt}\left(n_\sigma a^3\right) = 0,~~~~~~~~~n_\sigma(t) \equiv \frac{\rho_\sigma(t)}{m_\sigma(t)}.
\end{align}
Thus the comoving number density, $n_\sigma(t) a^3(t) \sim a^3 m_\sigma \left<\delta\sigma^2\right>$, is conserved.

%%%%%%%%
\section{Evolution of a scalar field with a non-renormalizable, temperature-dependent potential}
\label{app:nonren}

So far, we have considered a scalar field with a mass depending on the field value or temperature. In this section, we instead study the case where temperature-dependent non-renormalizable terms play a role. As a toy model, let us consider the potential
\begin{align}
    V = - \frac{1}{2} c H^2 \sigma^2 + \frac{T^2}{n M^{n-2}} \sigma^n,
\end{align}
where $c$ is a positive constant of ${\cal O}(1)$, $M$ denotes the cut-off scale of the non-renormalizable interaction, and $n$ is an integer larger than $2$, so that the second term represents a non-renormalizable operator. This kind of temperature-dependent coupling can arise from an operator of the form $\chi^2 |S|^{n}/M^{n-2}$ when $\chi$ particles are in thermal equilibrium. 

Following the analysis in Sec.~\ref{sec:Hub}, the temporary minimum $\sigma_{\rm min}$ is given by
\begin{align}
\sigma_{\rm min} & = \left(\frac{ cH^2 M^{n-2}}{T^2}\right)^\frac{1}{n-2} \propto a^{-\frac{3(1+ w)}{n-2}+ \frac{2p}{n-2}},
\end{align}
where we have assumed $T \propto a^{-p}$ with $p>0$. For instance, in the radiation-dominated era we have $p=1$ if the effective relativistic degrees of freedom remain constant, while in the inflaton-matter dominated era we have $p = 3/8$ if the inflaton decays into radiation with a constant decay rate.

On the other hand, the oscillation amplitude $\delta \sigma$ scales as (\ref{eq:delta sigma}). Thus, we have
\begin{equation}
    \frac{\delta\sigma}{\sigma_{\rm min}}
    \propto
    a^{-\frac{3}{4}(1-w) +\frac{3(1+ w)}{n-2}- \frac{2p}{n-2}}.
\end{equation}
The exponent is negative if
\begin{equation}
    w < \frac{-18+3n+8p}{3(n+2)}
\end{equation}
for $p \leq 3$. If $p \geq 3$, the exponent is always negative for $w < 1$. For instance, during the inflaton-matter dominated era ($w=0$, $p=3/8$), the condition requires $n > 5$. In the radiation-dominated era ($w = 1/3$, $p = 1$), it reduces to $n > 6$. 

To prevent the scalar field from crossing the origin, the other operators must remain subdominant until the Hubble-induced mass becomes comparable to the physical mass.

%%%%%%%%%%%%%%% References %%%%%%%%%%%%%%%%
\bibliographystyle{apsrev4-1}
\bibliography{ref}
%%%%%%%%%%%%%%%%%%%%%%%%%%%%%%%%%%%%%%%%%%%

\end{document}